\documentclass[]{aa} 
\usepackage[makeroom]{cancel}
\usepackage{xspace}
\usepackage{xcolor}
\newcommand{\aZ}{$\alpha_z$\xspace}
\newcommand{\aAv}{$\alpha_\mathrm{av}$\xspace}
\newcommand{\aAvSqr}{$\alpha_\mathrm{av}^\mathrm{sqr}$\xspace}
\newcommand{\aAvAbs}{$\alpha_\mathrm{av}^\mathrm{abs}$\xspace}
\newcommand{\aAvCenter}{$\alpha_\mathrm{av}^\mathrm{center}$\xspace}
\newcommand{\aAvSpot}{$\alpha_\mathrm{av}^\mathrm{spot}$\xspace}
\newcommand{\aPeak}{$\alpha_\mathrm{Peak}$\xspace}
\newcommand{\Br}{$B_{r}$\xspace}
\newcommand{\Bt}{$B_{\theta}$\xspace}
\newcommand{\Bx}{$B_{x}$\xspace}
\newcommand{\By}{$B_{y}$\xspace}
\newcommand{\Bz}{$B_{z}$\xspace}

\newcommand{\AB}[1]{\textcolor{blue}{[AB: #1]}}

\usepackage{graphicx}
\usepackage{txfonts}
\usepackage{makecell}
\usepackage{rotating}
\usepackage{tikz}
\usepackage{verbatim}
\usepackage{eqnarray}
\usepackage{amsmath}
\usepackage{ulem}

\begin{document} 

    \title{Impact of spatially correlated fluctuations in sunspots on metrics related to magnetic twist}

\author{
C.~Baumgartner\inst{\ref{inst1}}
\and
A.~C.~Birch \inst{\ref{inst1}}
\and
H.~Schunker \inst{\ref{inst2}}
\and
R.H.~Cameron  \inst{\ref{inst1}}
\and
L.~Gizon \inst{\ref{inst1},\ref{inst3}}
}

\institute{
Max-Planck-Institut f\"{u}r Sonnensystemforschung, 37077  G\"{o}ttingen, Germany \label{inst1}\\
\email{baumgartner@mps.mpg.de}
\and
School of Information and Physical Sciences, The University of Newcastle, New South Wales, Australia 
\label{inst2}
\and
Georg-August-Universit\"{a}t G\"{o}ttingen, Institut f\"{u}r Astrophysik, Friedrich-Hund-Platz 1, 37077   G\"{o}ttingen, Germany\label{inst3}
}

\date{Received 18 February 2022; accepted 17 March 2022}

\abstract{
The twist of the magnetic field above a sunspot is an important quantity in solar physics. For example, magnetic twist plays a role in the initiation of flares and coronal mass ejections (CMEs). Various proxies for the twist above the photosphere have been found using models of uniformly twisted flux tubes, and are routinely computed from single photospheric vector magnetograms. One class of proxies is based on $\alpha_z$, the ratio of the vertical current to the vertical magnetic field. Another class of proxies is based on the so-called twist density, $q$, which depends on the ratio of the azimuthal field to the vertical field. However, the sensitivity of these proxies to temporal fluctuations of the magnetic field has not yet been well characterized.
}
{
We aim to determine the sensitivity of twist proxies to temporal fluctuations in the magnetic field as estimated from time-series of SDO/HMI vector magnetic field maps. 
}
{
To this end, we introduce a model of a sunspot with a peak vertical field of 2370~Gauss at the photosphere and a uniform twist density $q= -0.024$~Mm$^{-1}$.  We add realizations of the temporal fluctuations of the magnetic field that are consistent with SDO/HMI observations, including the spatial correlations. Using a Monte-Carlo approach, we determine the robustness of the different proxies to the temporal fluctuations.
}
{
The temporal fluctuations of the three components of the magnetic field are correlated for spatial separations up to 1.4~Mm (more than expected from the point spread function alone). The Monte-Carlo approach enables us to demonstrate that several proxies for the twist of the magnetic field are not biased in each of the individual magnetograms. The associated random errors on the proxies have standard deviations in the range between $0.002$ and $0.006$ Mm$^{-1}$, which is smaller by approximately one order of magnitude than the mean value of $q$.}
{}

   {}

\maketitle


\newpage

\newpage

\section{Introduction}

\noindent
The magnetic field in solar active regions is often modeled by coherent bundles of magnetic field lines, so-called flux tubes. {The magnetic helicity $H = \int_V \mathbf{A} \cdot \mathbf{B} ~dV$, where $\mathbf{A}$ is the magnetic vector potential and $\mathbf{B}$ is the magnetic field, can be used to describe the topological structure of flux tubes fully contained in a volume $V$ \citep[][]{Berger1984}.} The magnetic helicity of a single flux tube has two components: writhe, which measures the deformation of its axis, and twist. If one imagines the magnetic field as a straight ribbon with its two ends rotated in opposite directions, the twist $T$ measures how often the ribbon turns around its straight axis
\begin{equation}
\label{eq:twist}
T = qL,
\end{equation}
where $L$ is the length of the ribbon and $q$ is the twist density, which counts how often the ribbon fully turns per unit length. 

\noindent
Measurements of the  twist of magnetic field play an important role in many different areas of solar physics: The  twist distribution in the photosphere constrains models of the solar dynamo and magnetic flux emergence \citep[e.g.,][]{Gilman1999,Brandenburg2005,Pipin2013}. 
For example, the hemispheric helicity sign rule describes an observed latitudinal dependence of the twist with predominantly negative (counter-clockwise) or positive (clockwise) twist in the northern or southern hemisphere \citep[][]{Seehafer1990,Pevtsov1995,Longcope1998,Nandy2006}, which is a key ingredient that solar dynamo models should be able to reproduce \citep[][]{Charbonneau2020}.
The twist in the  magnetic field plays an essential part in the dynamics of the solar atmosphere; for example a highly twisted flux tube can become susceptible to kink instability, which leads to a deformation of the axis of the  flux
tube in exchange for its twist. This is a possible trigger mechanism for 
solar flares and CMEs \citep[e.g., ][]{Torok2003,Torok2004,Leka2005,Fan2005}. Furthermore, the observed twist of photospheric magnetic field is used as an input to inject twist into coronal magnetic field extrapolations \citep[e.g., ][]{Yeates2008,Wiegelmann2012}.

Various methods have been developed to measure the twist
density of the magnetic field  in active regions directly from individual photospheric observations. These methods either use the force-free parameter, $\alpha$, as a proxy for the twist density or try to fit the twist density directly. 

{\cite{Woltjer1958} shows that the force-free parameter $\alpha$ in a closed system corresponds to the helicity content of a linear force-free field structure in its lowest attainable energy state. 
Therefore, $\alpha$ is used in observations as a proxy for the 
helicity of the magnetic field \citep{Pevtsov2014}.}

As we have observations at only one height in the photosphere from instruments like HMI, we cannot measure $\alpha$ directly. We are limited to calculating the vertical current density $J_z$ and consequently $\alpha_z = J_z/B_z$ at one height, where \Bz~is the vertical field strength. {\cite{Burnette2004} studied 34 active regions and show that a two-dimensional spatial average of \aZ over an active region is correlated with the $\alpha$ value corresponding to the three-dimensional linear-force-free extrapolation with the best least-squares fit to the observed field. 
Therefore,} spatial averages of \aZ are often used to characterize the helicity and twist of the magnetic field in active regions or individual sunspots \citep[e.g.,][]{Longcope1998, Hagino2004}. \cite{Leka2005} chose a single peak value of \aZ close to center the of a sunspot  (\aPeak) to characterize the twist in the magnetic field. This is because in simple models \aZ 
only relates directly to the twist density  at the axis of a flux tube ($\alpha_\mathrm{Peak} = 2q$).

\cite{Nandy2008} suggested a method to infer the twist density of the magnetic field $q$ that avoids the force-free assumption.
{These authors assume that the magnetic field in a sunspot can be approximated by a monolithic vertical flux tube with a constant twist density. They then use a least-squares fitting approach to fit the observations to this reference model to obtain the twist density.} 

Various tests of these methods have been conducted. \cite{Leka1999a} and \cite{Leka1999b} used observations to compare methods of using moments of the distribution of \aZ, a global $\alpha$ from force-free extrapolations, and a fitting approach of the function $J_z = \alpha B_z$. These authors find quantitative agreement between these methods. They also assessed the influence of instrumental effects like spatial resolution and a limited field-of-view, and considered noise by restricting the \aZ measurements to certain noise thresholds. \cite{Leka2005} successfully retrieved the twist density using their \aPeak method on a model by \cite{Fan2004} in the absence of errors. \cite{Crouch2012} evaluated different least-squares fitting methods of the twist density from a model flux tube, and find that the inferred twist density can be significantly different depending on the model assumptions used for fitting, also in the absence of noise. \cite{Tiwari2009_noise} used a linear force-free magnetic field model to test the effect of random polarimetric noise on estimates of the global $\alpha$ value of the synthetic field structure. 
{They find that noise does not influence the sign of $\alpha$ and the global twist can be measured accurately.} 

To interpret twist density measurements from a single observation, we need to understand how these measurements are affected by temporal variations of the magnetic field. HMI observes the magnetic field in the photosphere where the force-free assumption is thought to be violated \citep[][]{Gary2001}. 
{Temporal fluctuations of the magnetic field may arise in such an environment; for example} a twisted magnetic field structure can be distorted by its surrounding plasma flows. We need to model the fluctuations of the magnetic field in a sunspot from SDO/HMI observations to characterize the sensitivity of twist measurements to these fluctuations.

We tested the robustness of existing methods to infer the twist of the  magnetic
field under temporal variations of the magnetic field from SDO/HMI observations. We modeled the well-established leading sunspot of active region NOAA 11072 (observed by SDO/HMI at 2010.05.25 03:00:00 TAI) with the semi-empirical sunspot model by \cite{Cameron2011} with added uniform twist. We studied the spatial covariance of the  temporal fluctuations of the magnetic field and created a model based on our findings. We tested the robustness of methods to measure twist using Monte-Carlo simulations of the sunspot and fluctuation model.

In section~\ref{sec:observation} we present vector magnetic field observations of the reference sunspot in active region NOAA 11072. Sections~\ref{sec:noise_model} and~\ref{sec:sunspot_model} describe the  fluctuation and sunspot model, respectively. In section~\ref{sec:Twist_Calculation_Methods} we present a summary of the twist measurement methods that we test in this paper as well as their implementation. We then qualitatively compare our sunspot model to the SDO/HMI observations of the reference sunspot in section~\ref{sec:model_vs_observation}. In section~\ref{sec:mc_sim} we present Monte-Carlo simulations to test how the twist measurement methods fare under the influence of temporally fluctuating magnetic field.

\section{SDO/HMI vector magnetogram observations of our reference sunspot in active region NOAA 11072}
\label{sec:observation}

\begin{figure}
\centering
\includegraphics[scale=0.97]{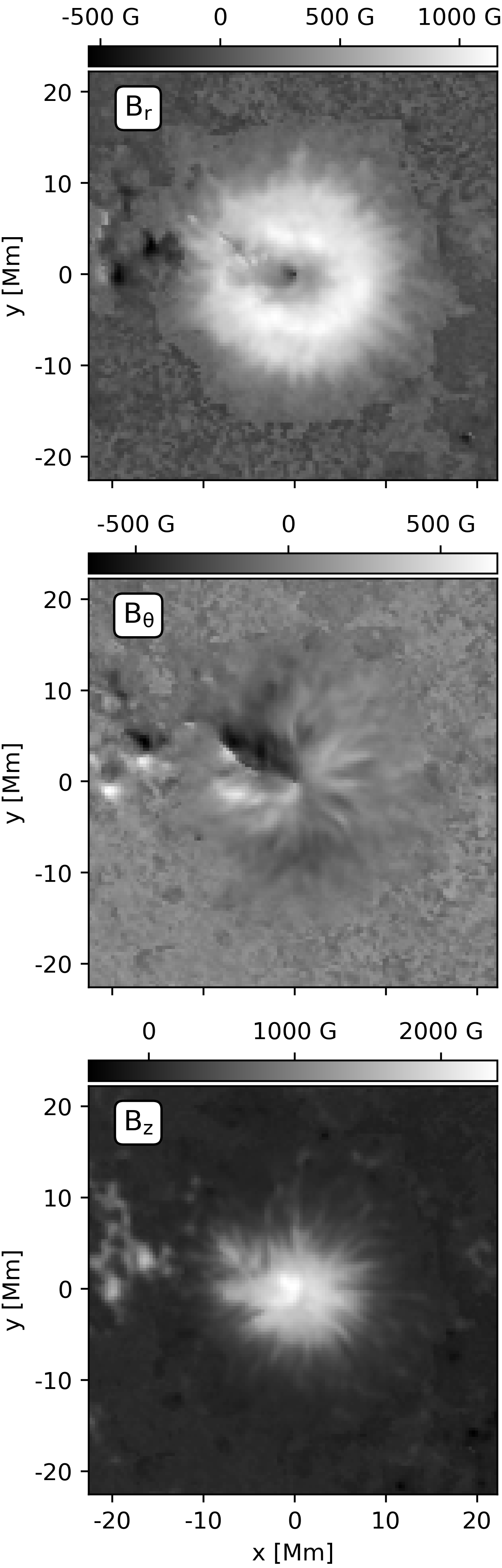}
\caption{SDO/HMI vector magnetogram of the leading
sunspot of active region NOAA 11072 (2010.05.25 03:00:00 TAI) in cylindrical coordinates. $B_r$ (top), $B_\theta$ (middle), and $B_z$ (bottom) show the radial, azimuthal, and vertical component of the magnetic field.}
\label{fig:cylindrical_plot}
\end{figure}

In this section, we present a sunspot observed by SDO/HMI that we selected as a reference for our sunspot model. The reference sunspot should closely resemble the assumption that its magnetic field structure could be described as a monolithic uniformly twisted flux tube. Therefore, we looked for sunspots that are well established, roughly circular, and under little influence from other strong magnetic field in its vicinity. We model the sunspot with uniform twist to test various methods to measure the twist under temporal fluctuations of the magnetic field.

We chose the leading sunspot of active region NOAA 11072 (2010.05.25 03:00:00 TAI), which was located about $30^\circ$ away from disk center at the Stonyhurst heliographic coordinates $27^\circ$~east and $13^\circ$~south. Figure~\ref{fig:cylindrical_plot} shows the Postel-projected sunspot transformed to local cylindrical coordinates from SDO/HMI vector magnetogram observations \citep[hmi.b\_720s,][]{Hoeksema2014}. $B_z$ is the component normal to the surface. $B_r$ and $B_\theta$ are located in a plane parallel to the surface. $B_r$ points radially away from the center of the  spot and $B_\theta$ is always perpendicular to $B_r$. See Appendix~\ref{sec:vector_transformation} for a detailed description of the coordinate systems and transformations used. 

\begin{figure}
\centering
\includegraphics[scale=0.62]{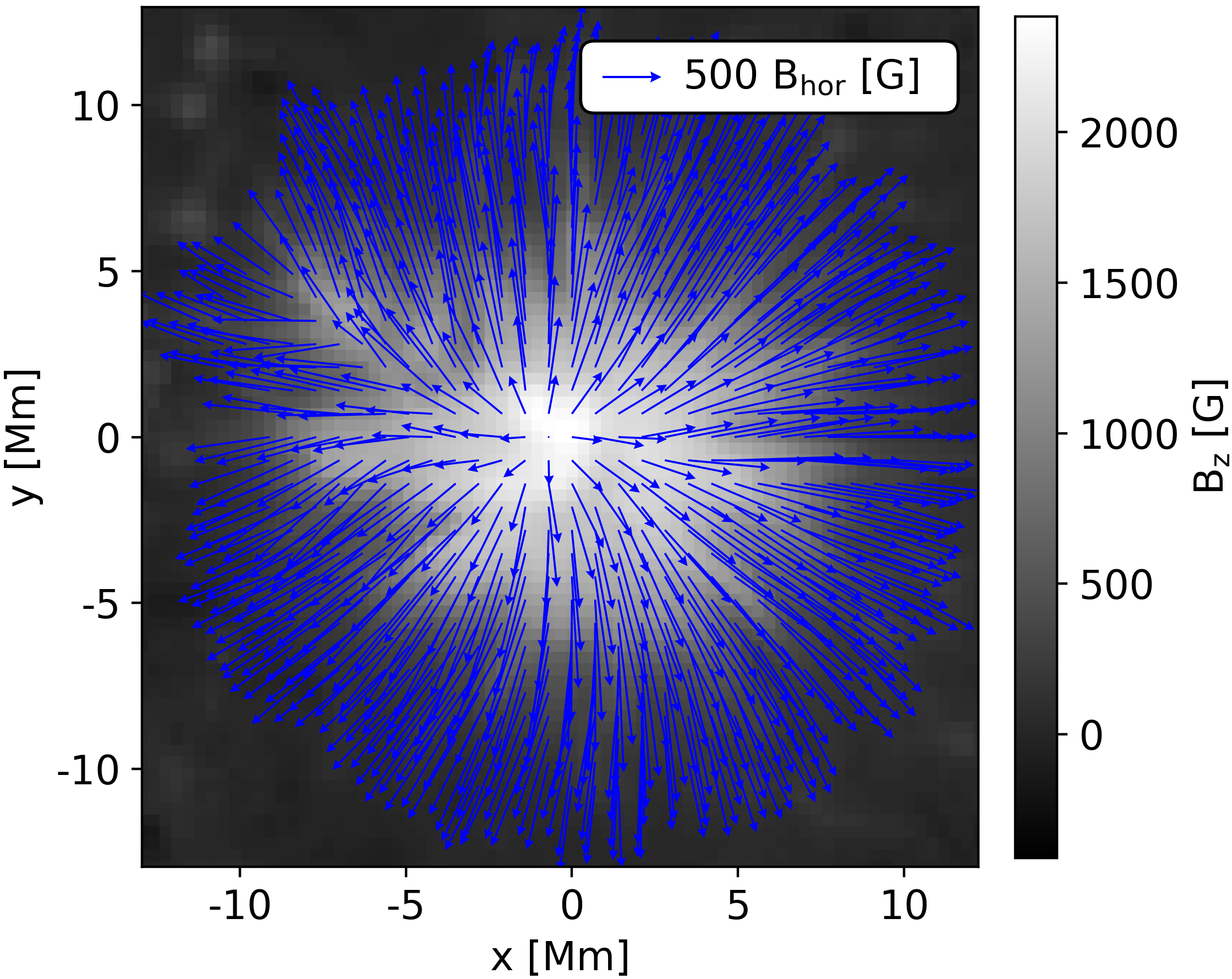}
\caption{SDO/HMI vector magnetogram of the  leading sunspot
of active region NOAA 11072 (2010.05.25 03:00:00 TAI) with the horizontal field $B_\mathrm{hor}$ plotted as arrows on top of the vertical vector component $B_z$.}
\label{fig:quiver}
\end{figure}

The sunspots we considered have a dominant radial $B_r$ and weak azimuthal $B_\theta$ component, as shown for the example of the reference sunspot in Fig.~\ref{fig:quiver}. This is a known characteristic of sunspots \citep[][]{Borrero2011_Sunspot}. We noticed that the azimuthal component ---which carries information about the handedness of the twist in a uniformly twisted flux tube--- shows that the sunspot has regions with opposite sign of twist.

\section{Estimating the spatial covariance of magnetic field fluctuations from the observations}
\label{sec:noise_model}

\begin{figure*}
\centering
\includegraphics[scale=0.35]{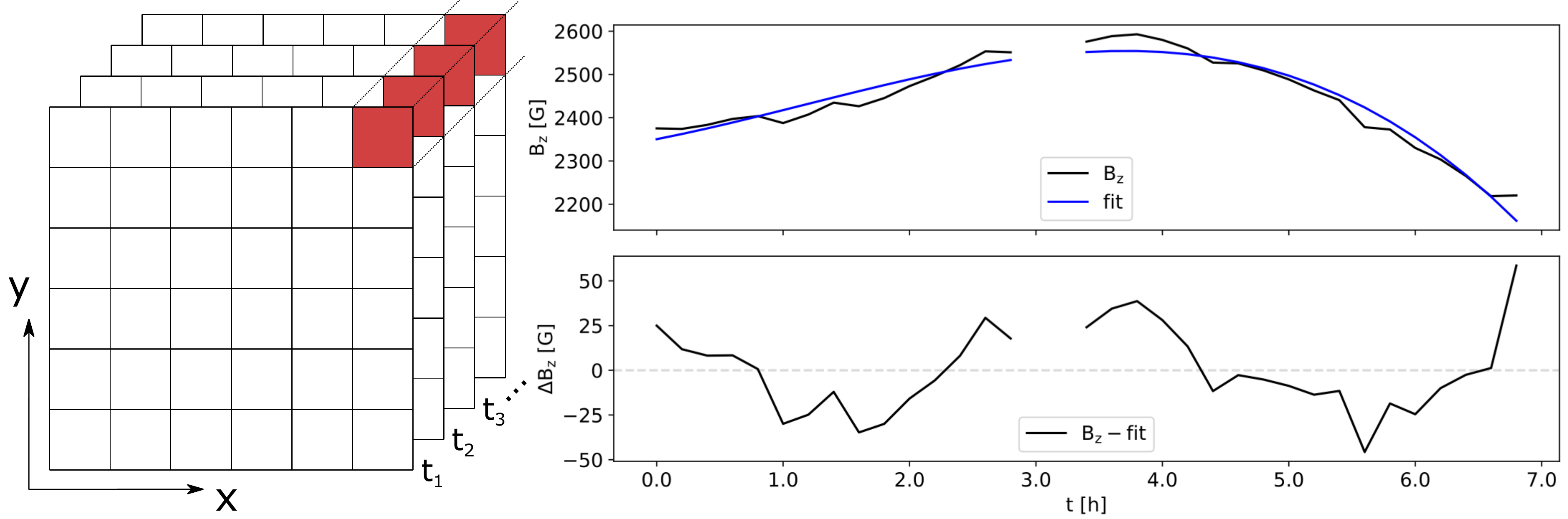}
\caption{Sketch of the detrending process. A series of consecutive vector magnetograms (e.g., the $B_z$ component) labeled with different time-steps ($t_1$, $t_2$, $t_3$,...), is shown on the left. The  black line (data) in the top right panel represents the temporal evolution of one pixel in this time-series (marked in red in the left panel). The blue line is a third-order polynomial fit to the data. The bottom right panel displays the detrended time-series, which shows the residuals of the data with respect to the fit.}
\label{fig:detrend_plot}
\end{figure*}

We aimed to derive a model for the temporal fluctuations of the magnetic field in SDO/HMI vector magnetogram observations. To do so, we used an approximately seven-hour time-series of our reference spot (2010.05.25, 03:00:00 - 9:48:00 TAI) to look at each local Cartesian vector component ($B_x$,~$B_y$,~$B_z$) in Postel-projected maps individually. The observations have a cadence of 12 minutes.
$B_z$ is the vector component normal to the surface, \Bx and \By point from solar east to west and south to north, respectively. The time-frame was chosen so that the spot is stable. 

We tracked the proper motion of the sunspot by first calculating the flux-weighted centroid of $|B_z|$ within the sunspot in each observation. We defined the area for this calculation based on pixels with a value below 0.85 in normalized continuum maps of the sunspot.
We found that the sunspot moved approximately three pixels over this time-period almost linearly. We fit a line to the location of the centroid in the x- and y- directions, and used the fit to shift the centroids of each image to the same location.

\begin{figure}
\centering
\includegraphics[scale=0.949]{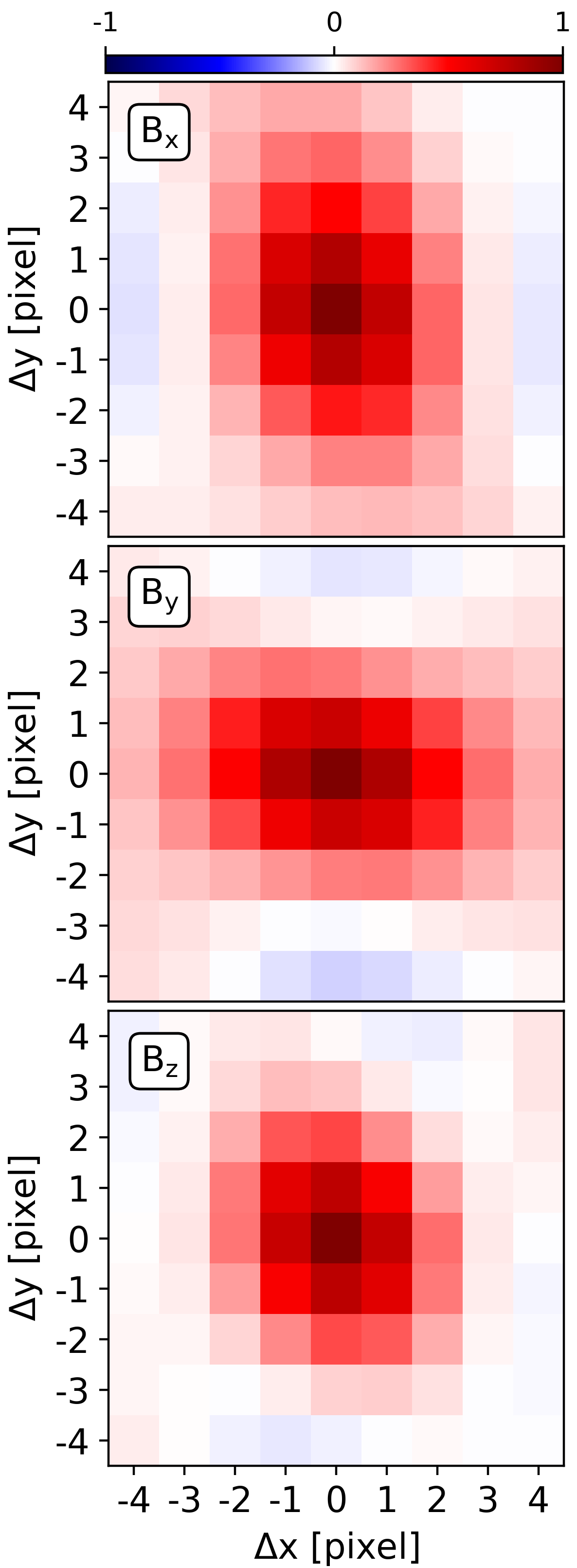}
\caption{Average spatial correlation of the detrended time-series of a pixel within the leading sunspot of active region NOAA 11072 (2010.05.25 03:00:00- 2010.05.25 9:38:00 TAI) relative to its neighbors for the magnetic field vector components $B_x$, $B_y$, and $B_z$.}
\label{fig:cor_plot}
\end{figure}

We then detrended the time-series of each pixel by fitting a third-order polynomial to the data and keeping the residuals (sketched in Fig.~\ref{fig:detrend_plot}) to remove any long-term trends. We calculated the spatial correlation of these detrended time-series using the Pearson correlation coefficent. Figure~\ref{fig:cor_plot} shows for each vector magnetic field component the average correlation within the sunspot of a pixel relative to its neighbors. We find that, on average, pixels in the observations are correlated with their neighbors up to 3-4 pixels away. 

 In Appendix~\ref{sec:correlation_test}, we consider whether the observed correlations were due to the Postel projection, the detrending method, the 12 and 24 hour periodicity of HMI caused by the satellites changing radial velocity relative to the Sun \citep[][]{Hoeksema2014}, or HMI's point spread function (PSF). We concluded that these correlations are caused by the PSF and the dynamic changes of the magnetic field over time. 

\begin{figure}
\centering
\includegraphics[scale=0.95]{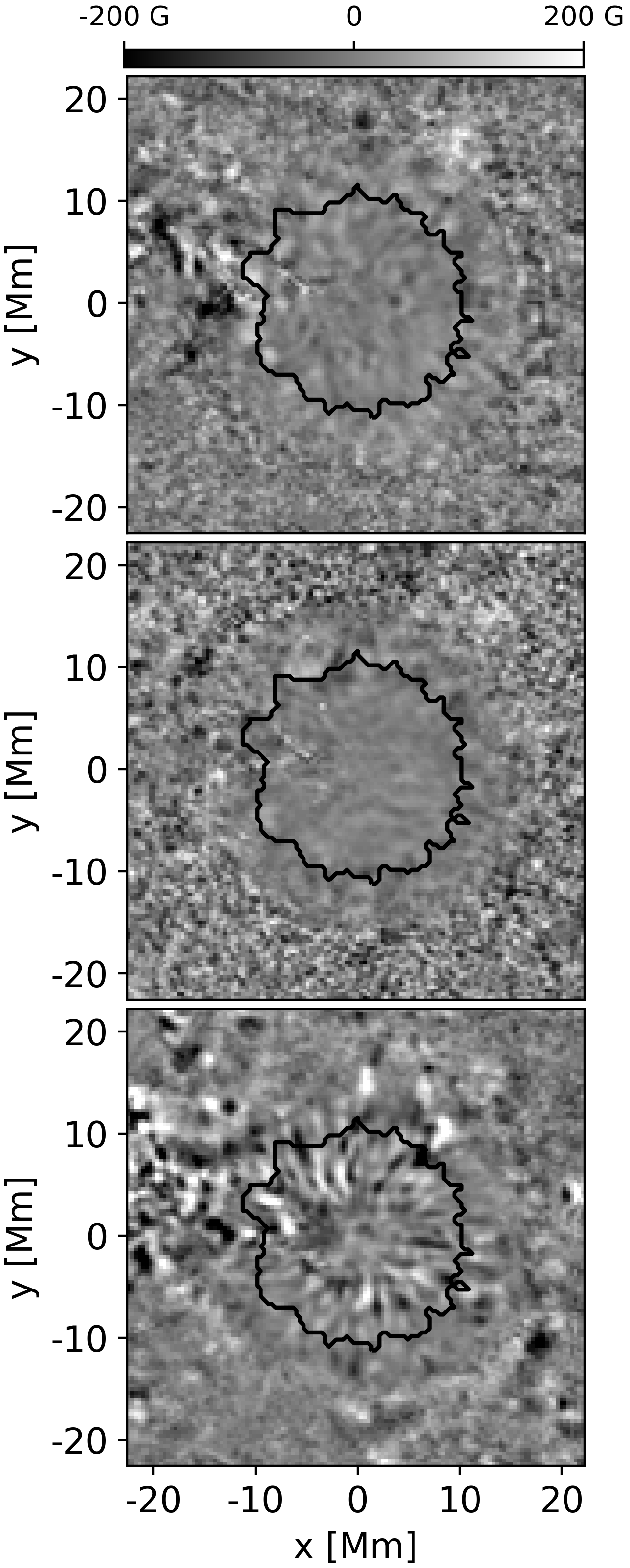}
\caption{Each panel represents one realization of magnetic field fluctuations for each vector component. All panels are plotted on the same scale in units of Gauss displayed by the color bar at the top. The black solid line represents the observed sunspot boundary.}
\label{fig:noise_plot}
\end{figure}

We incorporated the information about these time variations of the magnetic field into our model. Knowing that adjacent pixels are correlated but the strength of correlation depends on their position within the spot, we calculated the covariance matrix for all detrended time-series and for each vector component individually. We used the Cholesky decomposition \citep[][]{Haddad2009} of these covariance matrices to  create random correlated maps that reflect spatially correlated temporal fluctuations of the magnetic field.  Figure~\ref{fig:noise_plot} shows example realizations of such maps for each vector component. 

\section{A model for the reference sunspot}
 \label{sec:sunspot_model}
 
In this section we present the semi-empirical sunspot model developed by \citep[][]{Cameron2011} modified to represent a uniformly twisted field structure. \cite{Cameron2011} describes a three-dimensional magnetic field model of an axisymmetric sunspot with a radial \Br~ and vertical \Bz component. It lacks the azimuthal component \Bt, which is essential for creating twist. We added a \Bt~component that is only dependent on the radius without violating the requirement of $\nabla \cdot \mathbf{B} = 0$ in 3D space. We were only interested in the magnetic field structure on the photospheric level \mbox{(z = 0)} to model a sunspot observed by SDO/HMI. Specifically, the magnetic field at the photosphere is
\begin{eqnarray}
  \label{eq:model_Bz}
  B_z(r) &=& B_0 \exp \left[ -\left(\log_\mathrm{e}2\right)\left(\frac{r}{h_0}\right)^2\right],  \\
  \label{eq:model_Br} 
  B_r(r) &=& \frac{rB_z(r)}{8a_0\sqrt{1+b^2}}, \\
  \label{eq:model_Bt}
  B_\theta(r) &=& b B_r(r),
\end{eqnarray}
where $B_0$ is the magnetic field strength at the center of the  spot, $r$ is the distance from that  center, and $h_0$ defines the radius of the umbra--penumbra boundary. The parameter $a_0$ controls the inclination of the field and $b$ governs the amount of twist in the model. The model of 
{\cite{Cameron2011}  is designed so that the inclination of the magnetic field at the umbra--penumbra boundary is 45 degrees. Due to the additional azimuthal component \Bt and to keep the same inclination profile we adjusted the parameter controlling the inclination $a_0$ in accordance with the injected twist by multiplying it with $\sqrt{1+b^2}$.}

\noindent
The pixel scale of our model is the same as HMI's pixel scale of 0.5'', which corresponds to approximately 0.35 Mm at disk center.
{We fit the four free parameters ($B_0$,~$h_0$,~$a_0$,~$b$) to the reference spot.}
We use a least-squares fitting approach to best match the azimuthal averages of $B_z$, $B_r$, and $B_\theta$ around the flux-weighted
center of the reference sunspots  $|B_z|$. 

The information about the twist of the spot is stored in \Bt. As shown in Figs.~\ref{fig:cylindrical_plot} and~\ref{fig:quiver}, \Bt does not show an symmetric behavior about the center of the  spot and azimuthal averages do not represent the local twist present in the observation. We find that fitting only the positive or negative values of \Bt yields values of $0.125$ and $-0.175$ for the parameter $b$, respectively. As the magnetic field twist of the reference sunspot has a preference to be negative, we chose \mbox{$b = -0.15$}.


The parameters that we find to best describe the reference sunspot with uniform twist are \mbox{$B_0 = 2370~\mathrm{G}$},~\mbox{$h_0 = 5.2~\mathrm{Mm}$},~\mbox{$a_0 = 0.77~\mathrm{Mm}$}, and~\mbox{$b = -0.15$}.

\begin{figure*}
\centering
\includegraphics[scale=0.58]{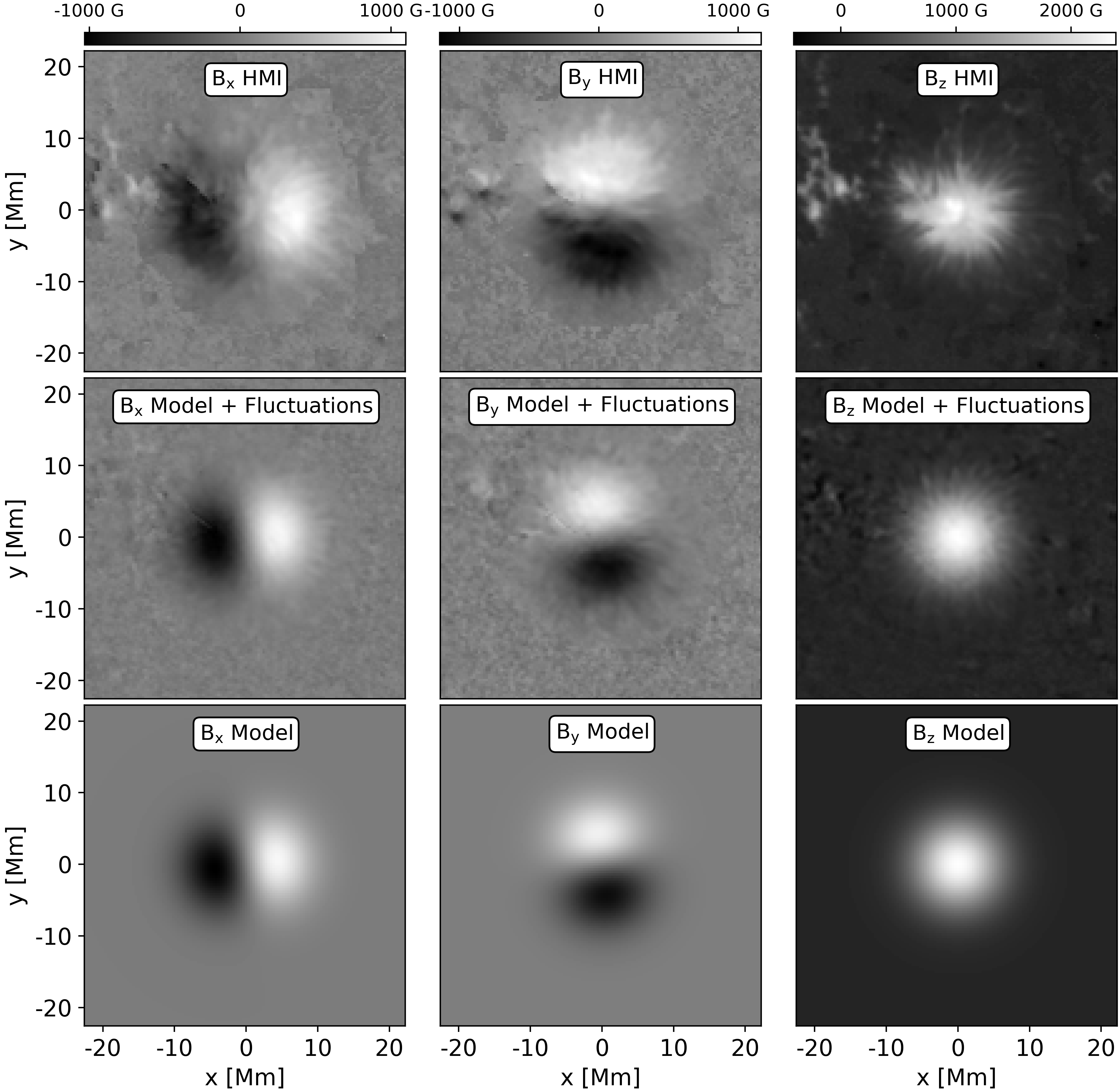}
\caption{Comparison of the vector components ($B_x$, $B_y$,
$B_z$) of the magnetic field. The first row shows the HMI observation of the  leading sunspot of active region NOAA 11072 (2010.05.25 03:00:00 TAI), which serves as the reference for our sunspot model. The second and third rows display the model with and without temporal fluctuations, respectively. Every column is plotted on the same scale in units of Gauss displayed by the color bar at the top of the column.}
\label{fig:model_plot}
\end{figure*}

\section{Summary and implementation of twist measurement methods}
\label{sec:Twist_Calculation_Methods}

Now we present various methods proposed in the literature to estimate the twist density directly from a single photospheric observation and describe their numerical implementation.

\subsection{The twist proxy $\alpha$}
\label{sec:alpha}

The force-free parameter~$\alpha$ can serve as a helicity proxy \citep[][]{Pevtsov2014} to estimate the twist in uniformly twisted flux tubes (see Appendix~\ref{sec:a_interpretation} for an interpretation of $\alpha$ in terms of twist). We lack information in photospheric observations of how the $B_x$ and $B_y$ components of the magnetic field change as a function of height (z-direction in a local Cartesian coordinate system) and we can only compute the vertical current density $J_z$ \citep[e.g.,][]{Pevtsov1994,Longcope1998}.

This vertical current density can be calculated from the force-free equation,
\begin{equation}
\mathbf{\nabla} \times \mathbf{B} = \mathbf{J} = \alpha \mathbf{B},
\label{eq:force_free_equation}
\end{equation}
with
\begin{equation}
  J_z = \frac{\partial B_y}{\partial x} - \frac{\partial B_x}{\partial y}.
 \label{eq:Jz_diff}
\end{equation}

Consequently we can compute the local twist proxy \aZ at a specific location with 
\begin{equation}
  \alpha_z =  \frac{J_z}{B_z}. 
  \label{eq:alpha}
\end{equation}
We note that $\alpha$ is a pseudo-scalar and the subscript $z$ denotes that it was derived only from the vertical field component and vertical current density. Positive (negative) values of \aZ correspond to right-handed (left-handed) magnetic field twist, respectively.

Numerically, we calculated the derivatives using the Savitzky-Golay filter \citep[][Appendix~\ref{sec:SavitzkyGolay}]{SavitzkyGolay1964} of cubic and quartic order and a stencil size of five pixels. We note that $J_z$ can also be calculated in integral form using Stokes' theorem (see Appendix~\ref{sec:Stokes}). The stencil size was chosen based on our own tests on how the stencil size impacts spacial averages of \aZ (see Appendix \ref{sec:boxsize_comparison}) and the results by \cite{Fursyak2018}.

\subsection{Average twist}

\cite{Pevtsov1995} used a single best fit value of $\alpha$ from linear force-free field extrapolations to characterize the twist for whole active regions.
\cite{Longcope1998} used an average $\langle \alpha_z \rangle$ for entire active regions, which can be calculated from photospheric observations without the need of any extrapolations:
\begin{equation}
  \alpha_{\mathrm{av}} = \langle \alpha_z \rangle = \left\langle \frac{J_z}{B_z} \right\rangle.
  \label{eq:alpha_best}
\end{equation}

\cite{Hagino2004} proposed two weighted averages of \aZ over a whole active region or spot to determine its twist: 
\begin{equation}
  \alpha_{\mathrm{av}}^\mathrm{abs} =  \frac{\sum J_z~\mathrm{sign}\left[B_z\right]}{\sum \left|B_z\right|}
  \label{eq:alpha_av_abs}
\end{equation}
and
\begin{equation}
  \alpha_{\mathrm{av}}^\mathrm{sqr} =  \frac{\sum J_z~B_z}{\sum B_z^2}.
  \label{eq:alpha_av_sqr}
\end{equation}
\aAvAbs and \aAvSqr are weighted by absolute and squared $B_z$, respectively, which is represented by the superscripts "$\mathrm{abs}$" and "$\mathrm{sqr}$".
These latter authors argue that these weighted averages have the advantage of putting less weight on weak field, especially close to the polarity inversion line, where singularities of \aZ are more likely to occur.

{The area over which \aZ is averaged depends on the area of interest that is studied. In section~\ref{sec:mc_sim} we investigate averages over a small central area of the spot, the umbra, and the whole spot.}

\subsection{Peak twist}
\label{sec:aPeak}

Under the assumption that a spot can be approximated by a monolithic uniformly twisted magnetic flux tube, \cite{Leka2005} show that the \aZ profile of this field structure has a peak directly at the center of the  spot (flux tube axis), which they name \aPeak. Based on the flux tube model by \cite{Gold1960}, \cite{Leka2005} demonstrate that \aPeak directly relates to the constant twist density $q$ of the field's structure ($\alpha_\mathrm{Peak} = 2q$).

In order to calculate \aPeak for a single sunspot, we estimated the location of the flux tube axis by computing the flux-weighted center  $|B_z|$  of the  sunspot. We calculated a map of \aZ values (Eq.~\ref{eq:alpha}) for each pixel within the spot and boxcar-smoothed this map to 2” as suggested by \cite{Leka2005}. We used the absolute values of this smoothed map to detect the \aZ-peak closest to the estimated flux tube axis. \aPeak is then the signed and smoothed \aZ value at the location of the  peak.

\subsection{Twist density}
\label{sec:Winding_Rate}

\cite{Nandy2008} proposed to fit the twist density $q$ to quantify the magnetic twist in a single spot.
Again, under the assumption that the magnetic field in a sunspot resembles a uniformly twisted flux tube, $q$ can be measured by fitting the slope of the equation
\begin{equation}
\frac{B_\theta}{B_z}  = qr ~+d,
\label{eq:winding_rate}
\end{equation}
where $r$ is the distance from the flux tube axis, and \Bt~ and \Bz~ are the magnetic field in azimuthal direction and along the axis of the  tube, respectively. The axis of the  tube is estimated by calculating the flux-weighted center of $|B_z|$ of the sunspot. We note that \cite{Nandy2008} allowed a nonzero intercept $d$ to occur in their Fig.~2. This violates the assumption of a vertical uniformly twisted flux tube, where the fitted function is expected to go through zero at $r=0$, which is equivalent to a vanishing \Bt~ at the axis of the 
flux tube. 
A physical interpretation of this intercept is not clear to us.

This method was carried out in a cylindrical coordinate system ($B_r$,~$B_\theta$,~$B_z$, see Appendix~\ref{sec:vector_transformation}). We fit the ratio $B_\theta$ over $B_z$ for each pixel as a function of the distance between the pixel and the estimated flux tube axis $r$. The resulting slope corresponds to the twist density $q$. We tested this method by both allowing an intercept and by forcing the fit through the origin (d=0).

\section{Example sunspot model with correlated magnetic field fluctuations compared to HMI observations}
\label{sec:model_vs_observation}

In this section we qualitatively compare the magnetic field and its twist between our sunspot model with one realization of magnetic field fluctuations and the SDO/HMI observations of the reference sunspot in active region NOAA 11072.

\subsection{The vector magnetic field}

Figure~\ref{fig:model_plot} shows that visually the model spot with correlated magnetic field fluctuations resembles the HMI observation of the leading sunspot of active region NOAA 11072 (2010.05.25 03:00:00 TAI). It even exhibits a filamentary structure especially noticeable in $B_z$, which is not present in the model without fluctuations of the magnetic field. The fluctuation maps (Fig.~\ref{fig:noise_plot}) have smoother and weaker fluctuations within the sunspot, and stronger, more chaotic fluctuations in the quiet Sun, as one would expect from the observations.

\subsection{The magnetic field's twist}

\begin{figure}
\centering
\includegraphics[scale=0.85]{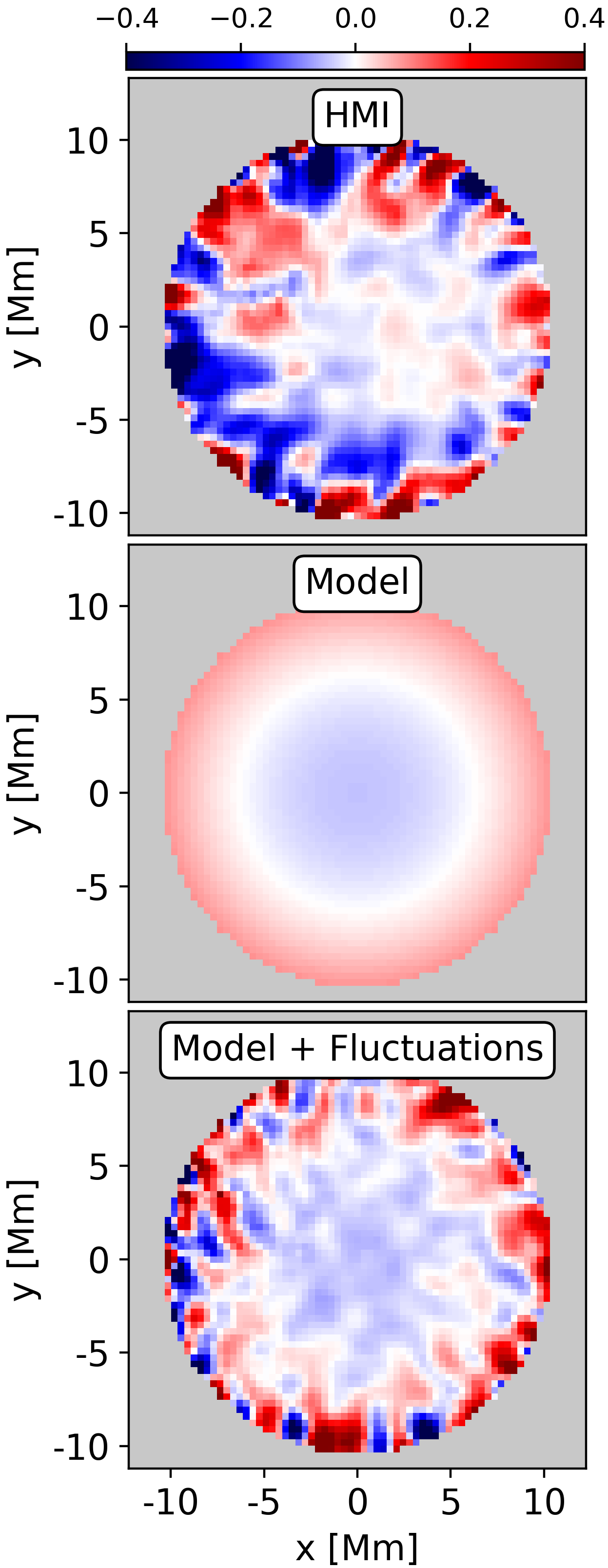}
\caption{Comparison of \aZ maps measured from the original HMI observation (top), and from the model without (middle) and with (bottom) fluctuations of the magnetic field. The color bar at the top shows the \aZ values in $\mathrm{Mm}^{-1}$.}
\label{fig:twist_examples} 
\end{figure}

Figure~\ref{fig:twist_examples} compares \aZ-profiles of the original SDO/HMI observation of the reference sunspot and our model with and without one realization of temporal fluctuations of the magnetic field.
Our sunspot model {without fluctuations} describes a uniformly left-handedly twisted magnetic field structure. 
{The \aZ-profile is azimuthally symmetric and \aZ increases with distance from the centers of the  spots. The sign of \aZ changes in the penumbra, which signals the presence of return currents (see Appendix~\ref{sec:a_interpretation}).}  

The reference sunspot has a more complicated structure than the model without temporal fluctuations. Even in the center of the spot, areas of opposite sign of \aZ exist. Towards the penumbra, positive values of \aZ become more frequent and one could assume a ring of return currents similar to the model. 
After applying magnetic field fluctuations to the model, a similar structure of the \aZ pattern compared to the observations develops.

{Our definition of fluctuations includes dynamic variations of the magnetic field}. Correlated changes in the direction of magnetic field  in adjacent pixels can produce spatially coherent changes in twist and its sign in our model. Sunspot observations typically show a strong radial field component and exhibit only weak twist (i.e. a weak azimuthal field component, $B_\theta << B_r$). {A source of fluctuations of the magnetic field is the forced environment of the photosphere, where the magnetic field can be buffeted by plasma flows, which can cause sign changes of the real twist.} Such an effect is expected to be stronger where the magnetic field strength is weaker ---for example in the penumbral parts of the sunspot--- where we see the strongest variations of \aZ within sunspots. Also, interactions with magnetic flux surrounding a sunspot could cause deviations from a uniformly twisted field structure. Complex patterns of \aZ and magnetic field twist even within the umbra of sunspots have been described in the literature \citep[e.g.,][]{Pevtsov1994,SocasNavarro2005,Su2009}.

\begin{figure*}
\centering
\includegraphics[scale=0.65]{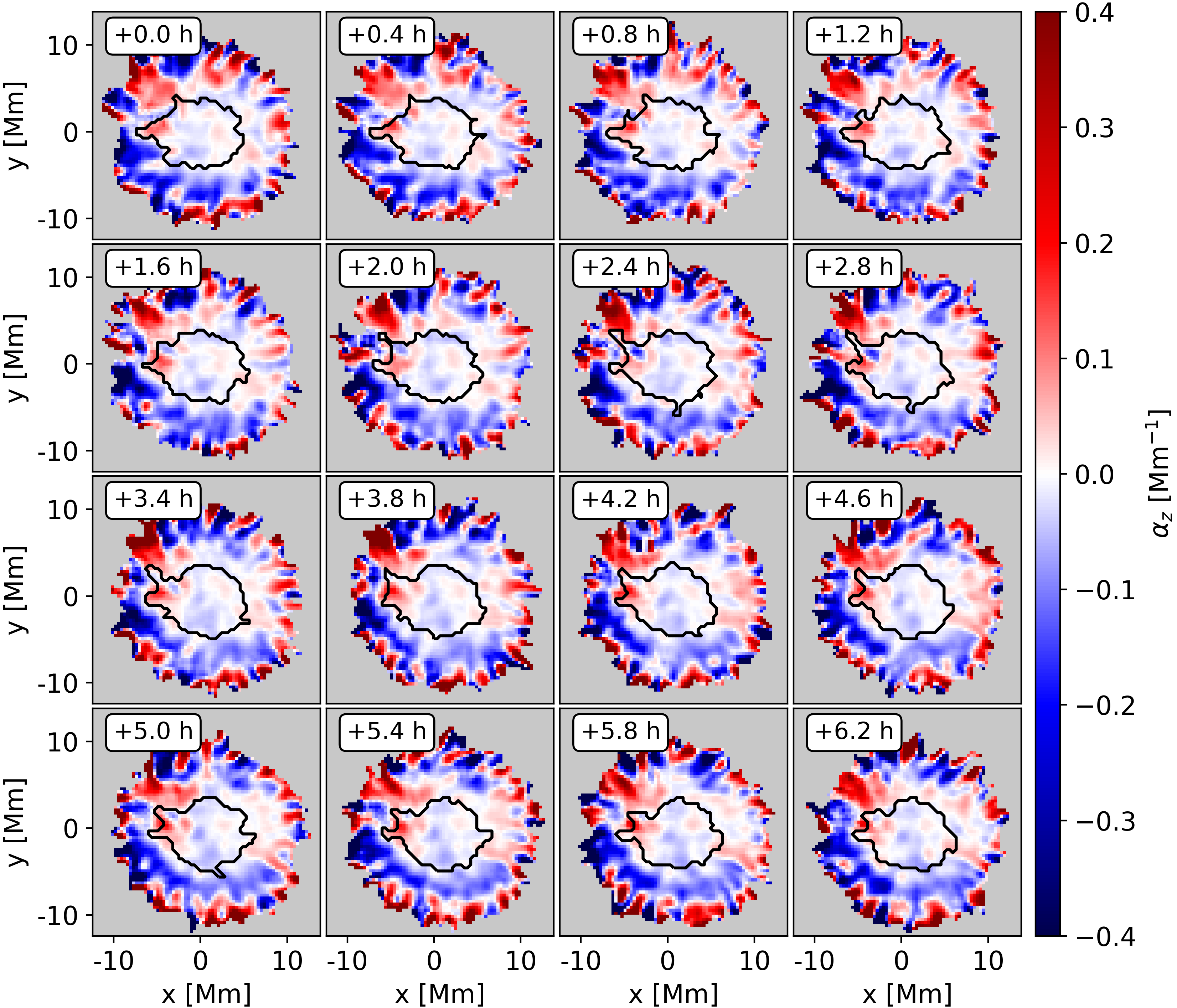}
\caption{Temporal evolution of \aZ in the leading sunspot of active region NOAA 11072. The time relative to the first observation (2010.05.25 03:00:00 TAI) is shown in hours in the top left of each panel. The black solid line represents the umbra--penumbra boundary.}
\label{fig:twist_observation}
\end{figure*}

\begin{figure*}
\centering
\includegraphics[scale=0.65]{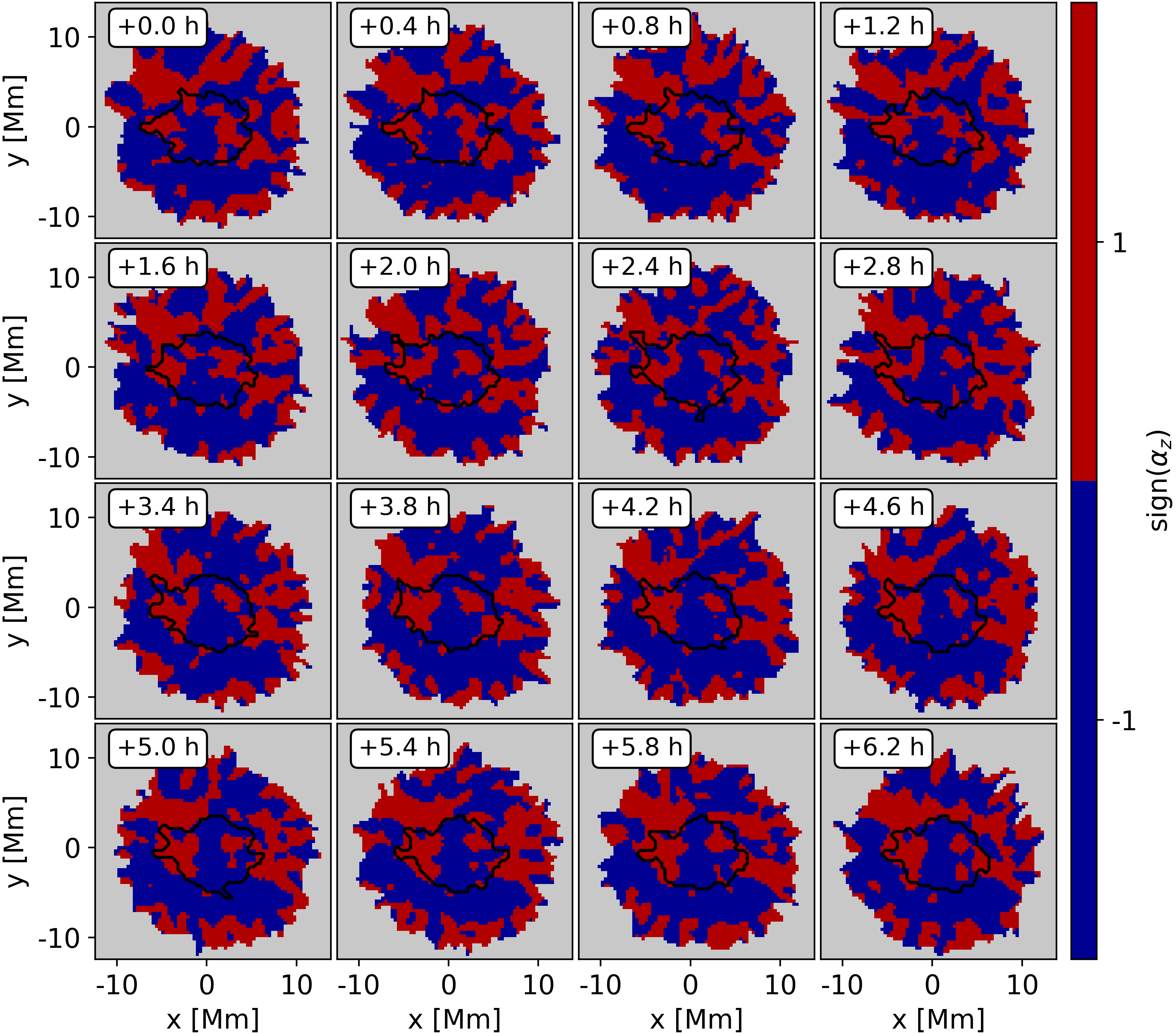}
\caption{Temporal evolution of the sign of \aZ in the leading sunspot of active region NOAA 11072. The time relative to the first observation (2010.05.25 03:00:00 TAI) is shown in hours in the top left of each panel. The black solid line represents the umbra--penumbra boundary.}
\label{fig:twist_sign_observation}
\end{figure*}

Figures~\ref{fig:twist_observation} and~\ref{fig:twist_sign_observation} show the temporal evolution of \aZ and its sign for the leading sunspot of active region NOAA 11072, respectively. In Fig.~\ref{fig:twist_observation} we find that in most parts of the umbra, \aZ is on the order of $10^{-2}~\mathrm{Mm^{-1}}$, while penumbral \aZ values are typically at least one order of magnitude larger, which is consistent with other \aZ-measurements in sunspots \citep[e.g.,][]{Tiwari2009,Wang2021}.

We find patches of \aZ with opposite signs throughout the reference sunspot, which is consistent with previous studies of sunspots \citep[e.g.,][]{Pevtsov1994,SocasNavarro2005,Su2009}. The shape of these patches are in agreement with findings by \cite{Su2009}, who describe a mesh-like pattern in the umbra and a thread-like pattern in the penumbra. Figure~\ref{fig:twist_sign_observation} shows that many of these patches persist over timescales of hours.

We find a similar distribution of \aZ values and patterns in the sunspot model with a realization of magnetic field fluctuations. The uncertainty on the mean \aZ (standard deviation of \aZ divided by the number of pixels considered) within the model sunspot with fluctuations is of the order of $10^{-3}~\mathrm{Mm}^{-1}$. This is of the same order as the uncertainty \cite{Leka1999a} and \cite{Leka1999b} measured in active regions observed with the Image Vector Magnetograph at Mees Solar Observatory.

Features in the pattern structure appear to be on a smaller scale in our model. \cite{Pevtsov1994} studied the  helical structure of the magnetic field of three active regions and estimated that the lifetime of such patches can exceed a day. {Our model only describes the spatial correlation of magnetic field fluctuations but does not address their temporal correlation. Therefore, these patches appear uncorrelated from one realization to  another}. Whether or not the temporal variations of the magnetic field are responsible for the twist and current patterns that we can observe in real sunspots cannot be deciphered yet from our model. To further investigate this question, one also has to consider the temporal correlation of the magnetic fluctuations.

\section{Sensitivity of twist measurements to correlated temporal fluctuations of the magnetic field}
\label{sec:mc_sim}

\begin{figure}
\centering
\includegraphics[scale=0.75]{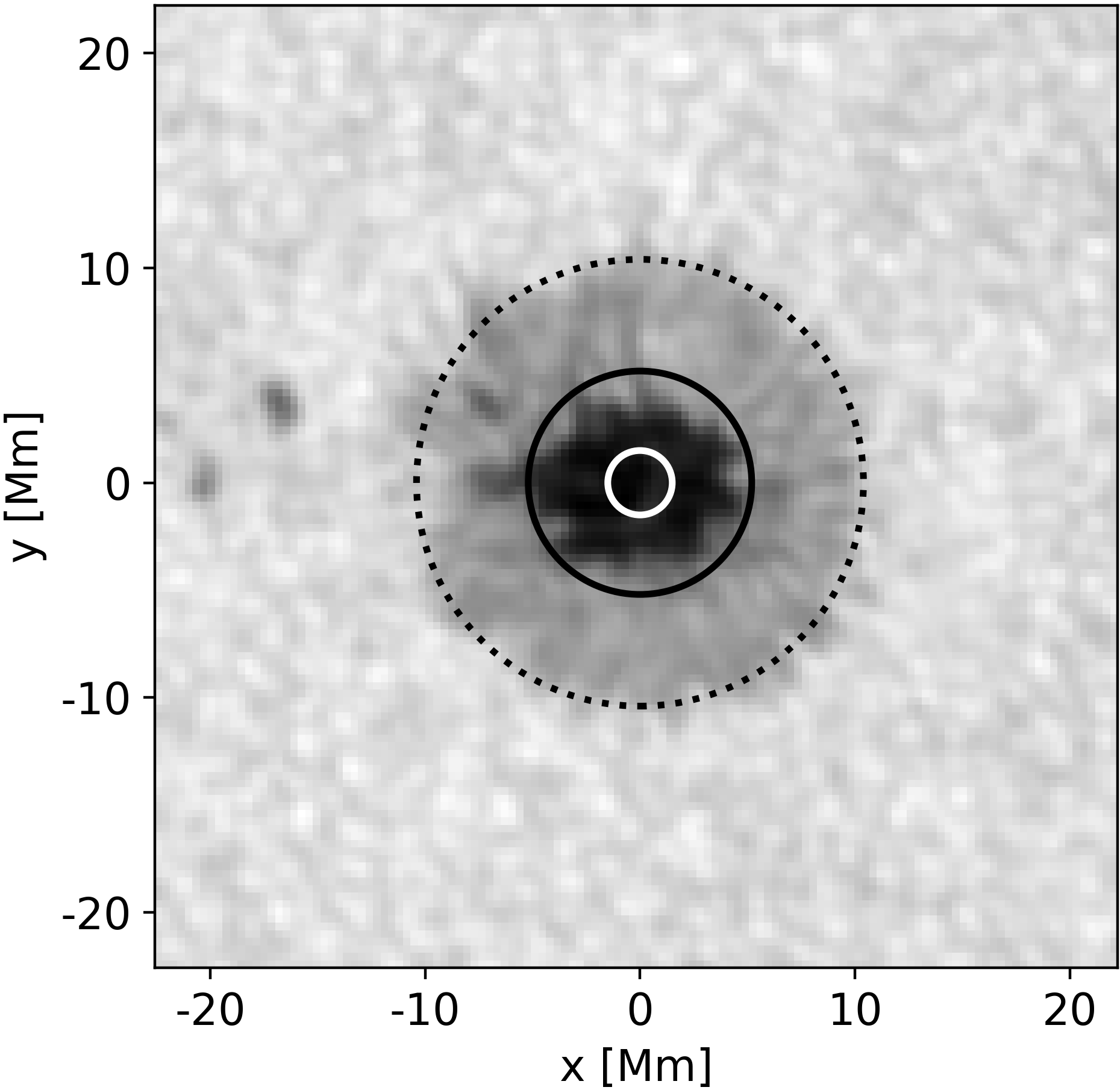}
\caption{HMI continuum image of the leading sunspot of active region NOAA 11072 (2010.05.25 03:00:00 TAI), which was used as a reference for the model presented in this work. The black solid line outlines the umbral area that was considered for most twist calculation methods. The white solid line and the black dotted line correspond to the areas that were used to get spatial averages of \aZ close to the center of the  spot (\aAvCenter) and over the whole spot (\aAvSpot), respectively.}
\label{fig:continuum}
\end{figure}

We used Monte-Carlo simulations to test the sensitivity of twist measurement methods described in section~\ref{sec:Twist_Calculation_Methods} to fluctuations of the magnetic field. We used our magnetic field fluctuation model (described in section~\ref{sec:noise_model}) in 10 000 realisations to create different fluctuation maps and superimposed them on the sunspot model (section~\ref{sec:sunspot_model}). We evaluated in each iteration the different twist proxies in the umbra (up to $r=h_0$). 
We also tested the robustness of the twist measurement methods based on the area  over which they are averaged. We evaluated \aAv in a small umbral area with a radius of 1.75 Mm from the center of the  spot (\aAvCenter, white circle in Fig.~\ref{fig:continuum}) and up to the penumbra--quiet Sun boundary (\aAvSpot, black dotted circle in Fig.~\ref{fig:continuum}, up to $r=2h_0$). Fluctuations of the magnetic field in our model can result in vertical field $B_z$ close to zero in the penumbra, which can create singularities when calculating $\alpha_z = J_z/B_z$. Therefore, we only measure \aZ for pixels that are above a $B_z$ threshold of 50~Gauss.
The analytically calculated reference values of \aZ and $q$ that represent the uniform twist of our model best (see Appendix~\ref{sec:a_interpretation}) are $\alpha_\mathrm{Peak}^\mathrm{ref} = 2q  \approx -0.048~\mathrm{Mm}^{-1}$.

\begin{figure*}
\centering
\includegraphics[scale=0.65]{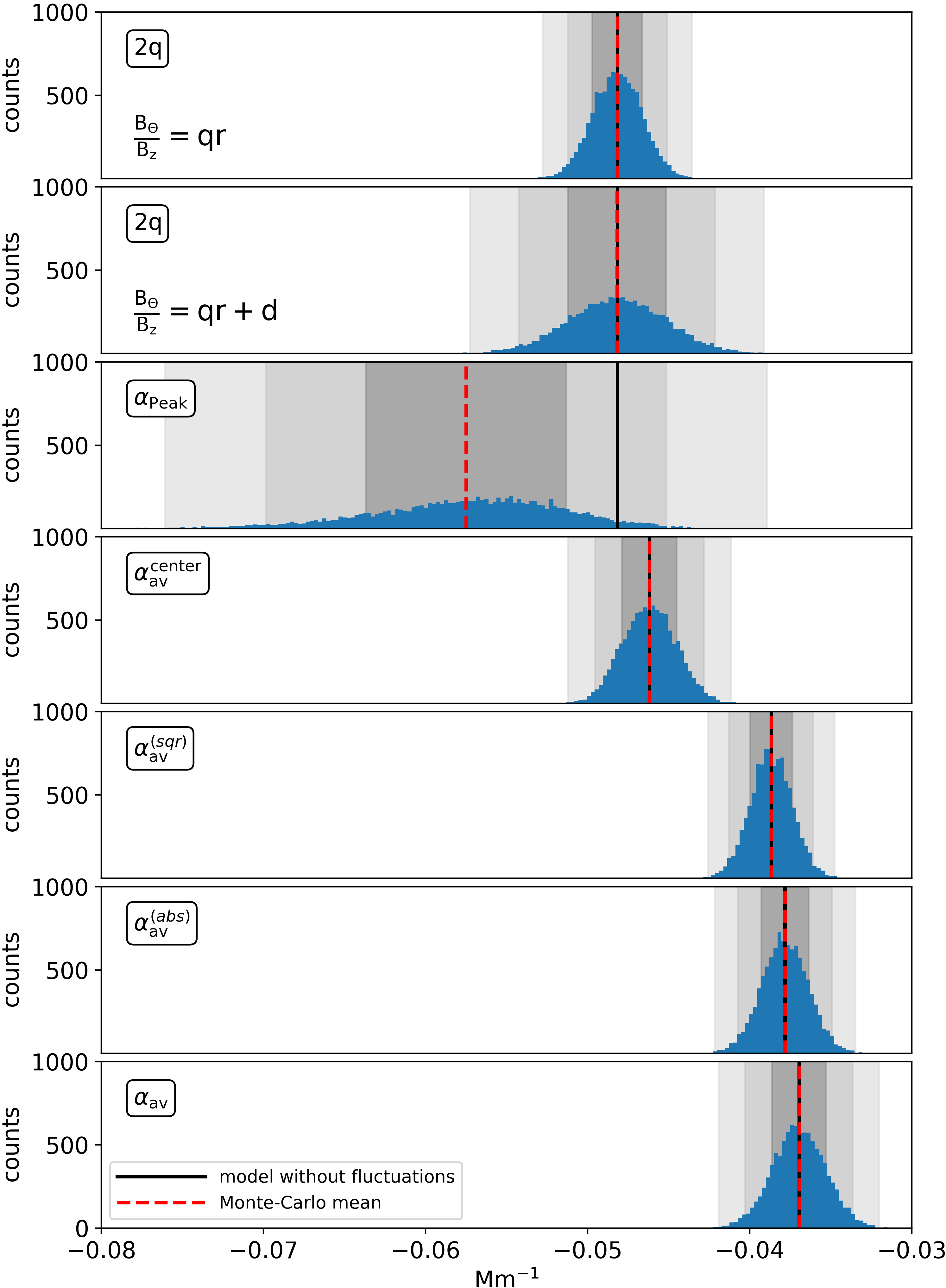}
\caption{Monte-Carlo simulation results for methods described in section~\ref{sec:Twist_Calculation_Methods}. The $q$ values have been multiplied by two in order to make them directly comparable to $\alpha_z$ (see Appendix~\ref{sec:a_interpretation}). The black line indicates the reference value of the model, which is expected from the model without any fluctuations. The red dashed line shows the mean value from the Monte-Carlo simulations. The gray shaded areas represent the range of 1, 2, and 3 $\sigma$ around the Monte-Carlo mean.}
\label{fig:twist_hist_dif}
\end{figure*}

\begin{figure*}
\centering
\includegraphics[scale=0.65]{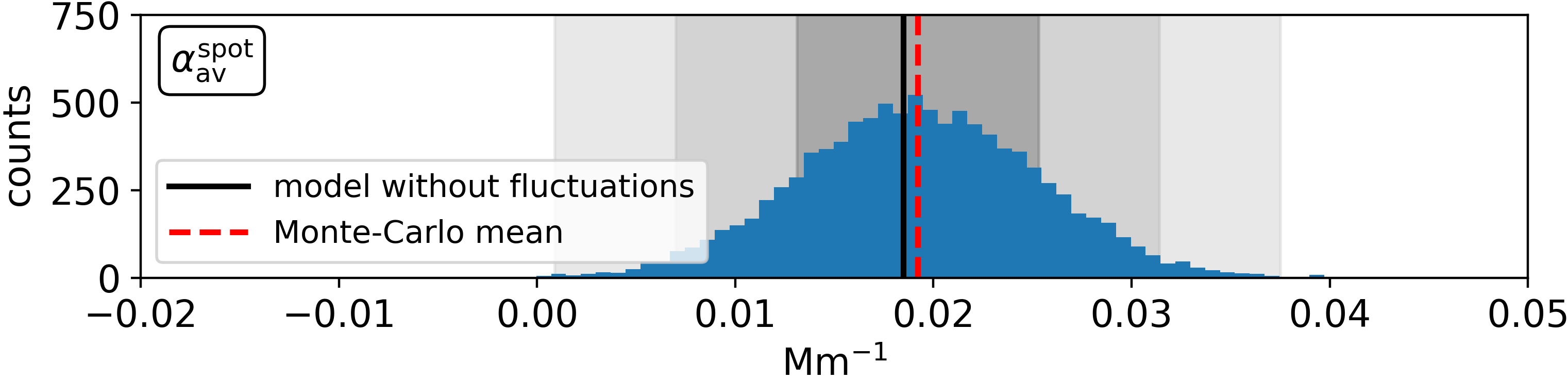}
\caption{Monte-Carlo simulation results when \aZ is spatially averaged over the whole sunspot. The solid black line indicates the reference value of the model that is expected from the model without any fluctuations. The red dashed line shows the mean value from the Monte-Carlo simulations. The gray shaded areas represent the range of 1, 2, and 3 $\sigma$ around the mean.}
\label{fig:twist_hist_spot_dif}
\end{figure*}

Figures~\ref{fig:twist_hist_dif} and~\ref{fig:twist_hist_spot_dif} show the distribution of the calculated twist proxies from the Monte-Carlo simulations for each method. Table~\ref{tab:results} compares the mean result and standard deviation from the Monte-Carlo simulations against the expected value from the model without fluctuations.
It is important to note that the "Model" values in Table~\ref{tab:results} are derived when a method is applied in the fluctuation-free model. The errors given in Table~\ref{tab:results} show the standard deviation of the Monte-Carlo simulations. 
We find that the expectation values of the averaging methods and the twist density fits are not biased by magnetic field fluctuations. 

{The averaging methods of \aZ show a large spread in their results, but have robust measurements under magnetic field fluctuations. These different spatial averages of \aZ can still be related to the twist density based on the azimuthal symmetric behavior of \aZ in our simple model. We derive 
$\alpha_z (r)= 2q\left[1-\left(\log_\mathrm{e}2\right)\left(\frac{r}{h_0}\right)^2\right]$ 
in Appendix~\ref{sec:a_interpretation}. We can calculate the average value of \aZ in a circular area with radius $r$ around the center of the spot:
\begin{equation}
    \label{eq:av_alpha_vs_q_integral}
    \begin{split}
        \langle \alpha_z \rangle_r &= \frac{1}{\pi r^2} \int_0^r \int_0^{2\pi} \alpha_z (\Tilde{r}) ~\Tilde{r}d\Tilde{r}d\theta =  
        \left[ 2-\left(\log_\mathrm{e}2\right) \left(\frac{r}{h_0}\right)^2\right] q.
    \end{split}
\end{equation}
Consequently, we find the following relation between the twist density $q$ and $\langle\alpha_z\rangle_r$:
\begin{equation}
\label{eq:av_alpha_vs_q}
    q = \frac{\langle\alpha_z\rangle_r}{2-\left(\log_\mathrm{e} 2\right) \left(r/h_0\right)^2}.
\end{equation}
This equation is consistent with the relation of $\alpha_{Peak}=2q$ at the center of the spot. It also describes the different spatial averages of \aZ that we measure, when the averaging radius was changed (see Fig.~\ref{fig:twist_hist_dif} and Fig.~\ref{fig:twist_hist_spot_dif}). We measure a larger $\langle \alpha_z \rangle$ when averaged over a small area at the  center of the spot ($\alpha_{\mathrm{av}}^\mathrm{center}$) compared to the average over the umbra ($\alpha_{\mathrm{av}}$). When the averaging radius becomes sufficiently large (e.g., $\alpha_{\mathrm{av}}^\mathrm{spot}$), $\langle \alpha_z \rangle$ retrieves the opposite sign compared to \aZ in the center of the spot. 
}

The expectation value of the \aPeak method is noticeably biased, causing an overestimation of the twist density in the presence of magnetic field fluctuations. As this method characterizes the twist density with a single peak value closest to the center of the spot, it is likely to pick up any enhanced signal caused by the fluctuations of the magnetic field.

The twist density fit \citep[][]{Nandy2008} retrieves the analytical twist density value. We find a larger spread in the resulting values when no zero-intercept is forced. These results are expected, because the equations of our sunspot model  can be exactly reduced to the fitting equation by \cite{Nandy2008}. \cite{Crouch2012} shows that fitting techniques can be sensitive to small discrepancies between the fitting and reference model. Observations suggest that sunspots are more complicated and can have nonuniformly twisted field structures  \citep[e.g.,][]{SocasNavarro2005,Su2009}. Another complication for real sunspots is that this method requires the exact location of the  axis of a flux tube as a center for the coordinate transformation to cylindrical coordinates. This is a simple task in our model without fluctuations, because the central axis of the  model spot and its flux-weighted center fall into the same place by definition; even with magnetic field fluctuations, they are always located close to each other. In real sunspots, the axis of the spot does not have to be in the same place as its flux-weighted center. Also, a sunspot may not have an underlying uniformly twisted vertical field structure.

{In section~\ref{sec:sunspot_model} we chose the model parameter $b=-0.15$, which governs the twist in our model. We tested various values of b, ranging from untwisted field ($b=0$) to highly twisted field ($b=10$) and found that the amount of twist in the model does not influence the findings about the robustness of the twist measurements described in this section.}

\begin{table}[]
\caption{Comparison of the different twist calculation methods between the model reference without any fluctuations of the magnetic field (Model) and the Monte-Carlo simulations (MC Sim.). All values are given in $\mathrm{Mm}^{-1}$.}
\centering
\begin{tabular}{l r r}
\hline\hline
Method & Model & MC Sim.   \\
\hline
$2q~\left(\frac{B_\Theta}{B_z} = qr\right)$      & $-0.048$ & $-0.048 \pm 0.002$ \\
$2q~\left(\frac{B_\Theta}{B_z} = qr~+~d\right)$  & $-0.048$ & $-0.048 \pm 0.003$ \\
$\alpha_\mathrm{Peak}$                              & $-0.048$ & $-0.058 \pm 0.006$ \\
$\alpha_\mathrm{av}^\mathrm{center}$                & $-0.046$ & $-0.046 \pm 0.002$ \\
$\alpha_\mathrm{av}^\mathrm{sqr}$                   & $-0.039$ & $-0.039 \pm 0.001$ \\
$\alpha_\mathrm{av}^\mathrm{abs}$                   & $-0.038$ & $-0.038 \pm 0.001$ \\
$\alpha_\mathrm{av}$                                & $-0.037$ & $-0.037 \pm 0.001$ \\
$\alpha_\mathrm{av}^\mathrm{spot}$                  & $0.019$ & $0.019 \pm 0.006$ \\

\hline
\end{tabular}
\label{tab:results}
\end{table}


\section{Summary and Conclusion}

We derived a model for the spatially correlated fluctuations of the magnetic field in a sunspot based on HMI observations of the leading sunspot of the active region NOAA 11072 . We superposed realizations of the fluctuations on the magnetic field of the semi-empirical sunspot model described in \cite{Cameron2011} with added uniform twist. We carried out Monte-Carlo simulations 
to test the robustness of the different measures of the twist to the fluctuations. 

{{We considered measurement methods based on estimating  either the force-free parameter~$\alpha_z$ or the twist density of the magnetic field ~$q$. In the absence of fluctuations, and for the sunspot model used in this paper, the value of \aZ at the center of the sunspot is twice the twist parameter $q$ \citep[see][]{Leka2005}.}}
For our chosen twist profile, \aZ is not uniform and changes sign in the penumbra of the model spot. 

Including the spatially correlated temporal fluctuations of the magnetic field qualitatively reproduces features seen in vertical current density $J_z$ and \aZ observations. Patches with opposite sign of \aZ appear throughout the sunspot model at random locations from one realization to another. Although we do not consider temporal correlations of the fluctuations, we note 
that such features can persist for hours in observations \citep{Pevtsov1994}.

All measures except \aPeak have expectation values consistent with that of the model without fluctuations. The fluctuations do not introduce a bias.
The measures based on spatial averages of \aZ have different expectation values, because they average over different portions of the nonuniform \aZ profile. 
Due to the sign change of \aZ in the penumbra of our model, any spatial averages of \aZ that reach too far out from the center of the spot can have the opposite sign from $\alpha_z$ near the center of the spot. 
The expectation value of \aPeak is biased with respect to the model without fluctuations. The magnitude of the bias is related to the level of the magnetic field fluctuations. 

For most methods, the spread of results from the Monte-Carlo simulations is less than the spread in expectation values of the individual methods. Therefore, the choice of method is more significant than the impact of the magnetic field fluctuations on the measurements.

Our results are for the particular sunspot model given by Eqs. \ref{eq:model_Bz}-\ref{eq:model_Bt}. This is a particularly simple axisymmetric sunspot model.
For sunspots with a more complex structure, for example nonuniformly twisted sunspots, the applicability and meaning of the different measures needs to be carefully considered \citep[e.g.,][]{Crouch2012}. In this regard, we note that the SDO/HMI observations of the leading spot of active region NOAA~11072 had fine structure in \aZ which persisted for longer than 7 hours. This persistent fine structure suggests a more complicated underlying field structure than that of our uniformly twisted model.
The long-lived fine structure found in active region NOAA~11072 is consistent with previous studies of other sunspots \citep[e.g.,][]{Pevtsov1994,Su2009}.

{{
As was previously noted by \cite{Leka1999b}, a single parameter will not in general characterize the twist of a sunspot.
In this paper, we show that a range of different complementary parameters exists, all of which describe somewhat different aspects of the magnetic field line twist in sunspots. We show that most of these measures are robust to fluctuations of the field. }}

\begin{acknowledgements}
CB is a member of the International Max Planck
Research School (IMPRS) for Solar System Science at the University of
Göttingen. CB conducted the analysis, contributed to the interpretation of
the results, and wrote the manuscript. We thank Jesper Schou for helpful discussions. We thank Graham Barnes for useful comments on the manuscript. The HMI data used are courtesy
of NASA/SDO and the HMI Science Team. ACB, RHC and LG acknowledge partial support
from the European Research Council Synergy Grant WHOLE SUN \#810218.
The data were processed at the German Data Center for SDO, funded by the
German Aerospace Center under grant DLR50OL1701. This research made use of the
Astropy \citep[][]{astropy2013,astropy2018},  Matplotlib \citep[][]{matplotlib}, NumPy \citep[][]{numpy} and  SciPy \citep[][]{scipy} Python packages.

\end{acknowledgements}

\bibliographystyle{aa} 
\bibliography{bibliography.bib} 

\appendix

\section{Vector transformation}
\label{sec:vector_transformation}
\subsection{Local Cartesian coordinates} \label{sec:local_cartesian}
The hmi.b\_720s \citep{Hoeksema2014} provides the vector magnetic field in spherical coordinates aligned with the line of sight (LoS); it includes the absolute field strength $B$, the inclination angle ``inc'' with respect to the LoS, and the azimuth angle $\mathrm{``azi}$'' (whose ambiguity has to be resolved) measured in a plane perpendicular to the LoS. We study the magnetic field in a local Cartesian and cylindrical coordinate system, where \Bz~ is pointing radially outwards from the Sun. In order to transform the spherical coordinate system to a local Cartesian coordinate system, we followed the transformations described in \cite{Gary1990}. 

First, the spherical vector components were transformed to a Cartesian system, where $B_\zeta$ is aligned with the LoS, while $B_\xi$ and $B_\eta$ lay in the plane perpendicular to the LoS:  
\begin{eqnarray}
\label{eq:spherical_to_LoS_Cartesian}
B_\xi &=& -B\sin{\mathrm{(inc)}\sin{\mathrm{(azi)}}} \\
B_\eta &=& B\sin{\mathrm{(inc)}\cos{\mathrm{(azi)}}} \\
B_\zeta &=& B\cos{\mathrm{(inc)}}.
\end{eqnarray}
We then rotated the coordinate system (Eq.~\ref{eq:LoS_Cartesian_to_Local_Cartesian}) so that $B_\zeta$ points radially outward from the Sun and becomes $B_z$. Here, \Bx~ and \By~ point from solar east to west and south to north, respectively. The rotation matrix is 
\begin{equation}
\label{eq:LoS_Cartesian_to_Local_Cartesian}
\begin{bmatrix} B_x \\ B_y \\ B_z
\end{bmatrix}
= \begin{bmatrix}
a_{11}~ a_{12}~ a_{13} \\
a_{21}~ a_{22}~ a_{23} \\
a_{31}~ a_{32}~ a_{33}
\end{bmatrix}
\begin{bmatrix}
B_\xi \\ B_\eta \\ B_\zeta
\end{bmatrix}
,\end{equation}
with the transformation matrix coefficients $a_{ij}$: 
\begin{eqnarray*}
a_{11} &=& - \sin B_0\sin P\sin(L-L_0)+\cos P\cos(L-L_0) \\
a_{12} &=& + \sin B_0 \cos P \sin(L-L_0) + \sin P \cos(L-L_0) \\
a_{13} &=& - \cos B_0 \sin (L-L_0) \\
a_{21} &=& - \sin B \left[ \sin B_0 \sin P \cos(L-L_0) + \cos P \sin(L-L_0) \right] - \\
       & & \quad - \cos B \left[ \cos B_0 \sin P \right]\\
a_{22} &=& + \sin B \left[ \sin B_0 \cos P \cos(L-L_0) - \sin P \sin(L-L_0) \right] + \\
a_{23} &=& - \cos B_0 \sin B \cos(L-L_0) + \sin B_0 \cos B\\
a_{31} &=& + \cos B \left[ \sin B_0 \sin P \cos(L-L_0) + \cos P \sin(L-L_0) \right] - \\
       & & \quad - \sin B \left[ \cos B_0 \sin P \right] \\
              & &\quad + \cos B \left[ \cos B_0 \cos P \right] \\
a_{32} &=& - \cos B \left[ \sin B_0 \cos P \cos(L-L_0) + \sin P \sin(L-L_0) \right] + \\
       & & \quad + \sin B \left[ \cos B_0 \cos P \right] \\
a_{33} &=& + \cos B \cos B_0 \cos(L-L_0) + \sin B \sin B_0.
\label{eq:matrix_coefficients}
\end{eqnarray*}
Here, $L$ and $B$ describe the heliographic longitude and latitude of the individual pixels, 
while $L_0$ and $B_0$ are the longitude and latitude of the solar disk center, respectively, and $P$ is the solar position angle.

\subsection{Cylindrical coordinates}
\label{sec:local_cylindrical}

We  transformed from the local Cartesian coordinate system to a local cylindrical coordinate system by calculating 
\begin{equation}
\label{eq:Cartesian_Cylindrical}
\begin{aligned}
\begin{bmatrix} B_r \\ B_\theta \\ B_z
\end{bmatrix}
= \begin{bmatrix}
&\cos\theta~ &\sin\theta ~&0\\
&-\sin\theta~ &\cos\theta~&0\\
&0~ &0~ &1
\end{bmatrix}
\begin{bmatrix}
B_x \\ B_y \\ B_z
\end{bmatrix},
\end{aligned}
\end{equation}
with $\theta = \arctan2(y,x)$. The flux-weighted center of a spot is defined as the origin ($x = 0, y = 0$) for this transformation. In this coordinate system, $B_z$ is the component normal to the surface; $B_r$ and $B_\theta$ are located in a plane parallel to the surface; $B_r$ points radially away from the center of the spot; and $B_\theta$ is always perpendicular to $B_r$ and $B_z$.

\section{Possible causes for the measured spatial correlation of magnetic field fluctuations}
\label{sec:correlation_test}

We tested various effects that could introduce the spatial correlation of temporal fluctuations of neighboring pixels.
To assess the effect of the Postel projections on these correlations, we created artificial full-disk HMI maps. We filled pixels with random Gaussian white noise with a mean of zero and standard deviation of one. We created Postel-projected time-series of submaps at different locations on the solar disk based on the center to limb angle (CTL) of the  center of the submap. We followed the same detrending procedure as described in section~\ref{sec:noise_model}, but find no correlation of adjacent pixels either at disk center (CTL = $0^\circ$) or close to the limb (CTL = $60^\circ$). 

We compared different ways of detrending the data: different order of polynomials for fitting (order 3, 4, and 5), differences to previous data points, and running averages. All different detrending methods show similar correlations of adjacent pixels.
We find no relation of the detrended signals to the known 12- and 24-hour period systematic errors of HMI that are caused by the satellite's orbit and change in radial velocity relative to the Sun.

\begin{figure}
\centering
\includegraphics[scale=0.75]{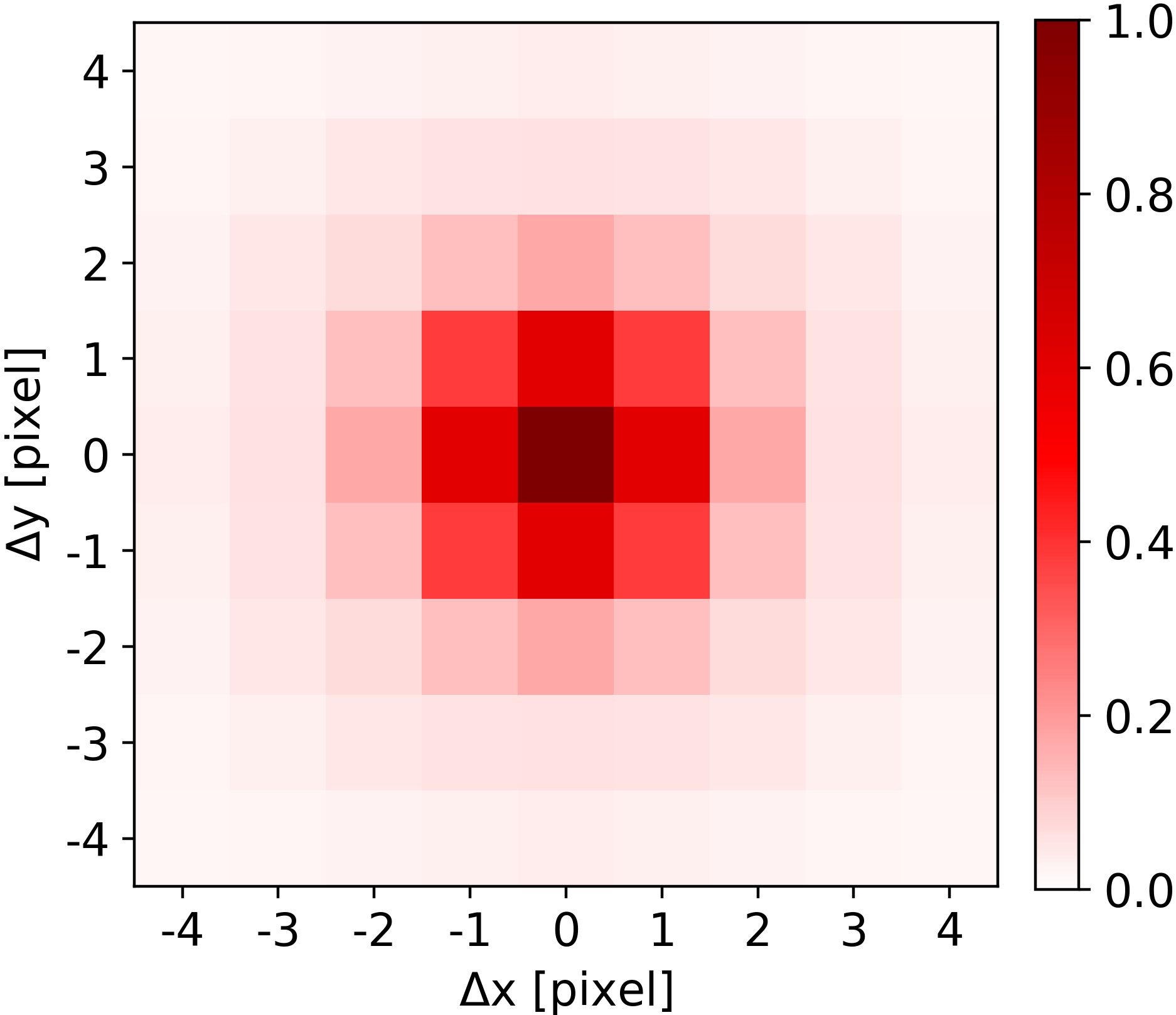}
\caption{Normalized PSF for HMI estimated by \cite{Yeo2014}.}
\label{fig:PSF_Yeo}
\end{figure}

\begin{figure}
\centering
\includegraphics[scale=0.60]{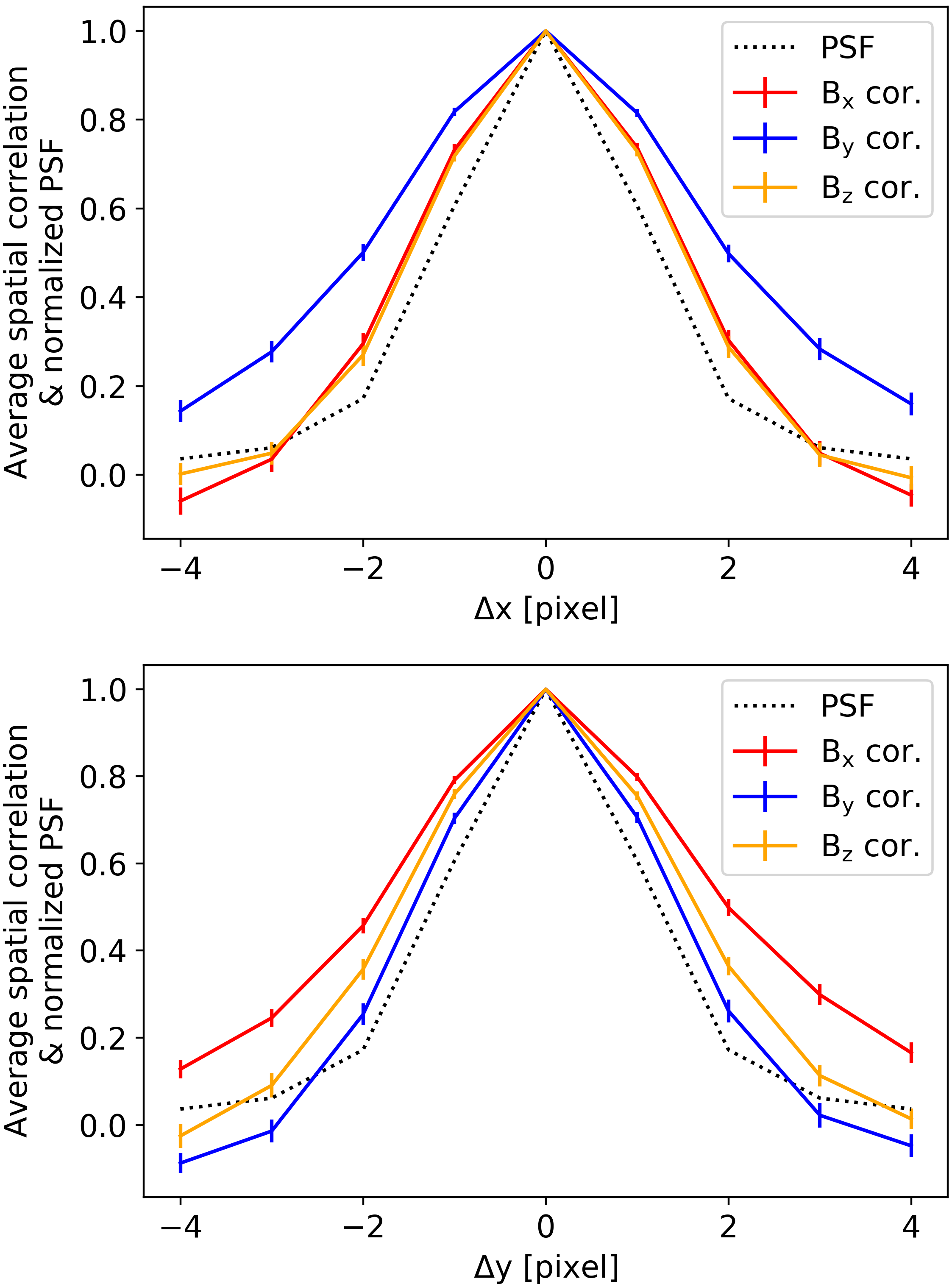}
\caption{Cuts through the center along the x-axis (top panel) and the y-axis (bottom panel) of the average spatial correlation of neighboring pixels (Fig.~\ref{fig:cor_plot}) and the normalized estimated PSF by \cite{Yeo2014} (Fig.~\ref{fig:PSF_Yeo}). The errorbars represent the standard error of the spatially averaged correlation.}
\label{fig:psf_cut}
\end{figure}

The PSF of HMI definitely contributes to the correlations of adjacent pixels. Figure~\ref{fig:PSF_Yeo} shows an estimate of the PSF of  HMI by \cite{Yeo2014}. Figure~\ref{fig:psf_cut} shows cuts along the x- and y-axis through the center of the normalized PSF, and the average correlations of adjacent pixels are shown in Fig.~\ref{fig:cor_plot}.  We find that the fluctuations of the vector magnetic field components are correlated up to spatial scales that are 30\% larger than that expected from the HMI PSF.
We suggest that these correlations are caused by the PSF and the dynamic changes of the magnetic field over time with respect to the underlying global field structure of the sunspot.

\section{Derivatives with different stencil sizes }
\label{sec:SavitzkyGolay}

We can evaluate derivatives with different stencil sizes using Savitzky-Golay filter \citep[][]{SavitzkyGolay1964} of cubic/quartic order for stencil sizes of 5 and 7 pixels or take a central differences approach for a stencil of 3 pixels:
\begin{equation}
\label{eq:sav_golay}
    \frac{\partial f(x)}{\partial x} = \frac{1}{N h} \sum_{i=-(S-1)/2}^{(S-1)/2} g(i)f(x+i),
\end{equation}
where $x$ describes the location of a pixel counted as integer, for which we want to calculate the derivative. $S$ is the stencil size, $N$ is a normalization factor, $h$ is the pixel scale, and $g$ is a weighting factor. The weighting and normalization factor for the different stencil sizes are listed in Table~\ref{tab:SavitzkyGolay}.

\begin{table}[]
\caption{Weighting factors g(i) and normalization factors $N$ for stencil sizes S of 7, 5, and 3 pixels for Eq.~\ref{eq:sav_golay}.}
\centering
\begin{tabular}{c | r r r}
\hline\hline
$i$   &       &   $g(i)$        \\
   &   $S=7$   &   $S=5$   &   $S=3$   \\

\hline
$-3$  & $22$    &   $0$   &   $0$   \\                             
$-2$  & $-67$   &   $1$   &   $0$   \\
$-1$  & $-58$   &   $-8$  &   $-1$  \\
$0$   & $0$     &   $0$   &   $0$   \\
$1$   & $58$    &   $8$   &   $1$   \\
$2$   & $67$    &   $-1$  &   $0$   \\
$3$   & $-22$   &   $0$   &   $0$   \\
\hline\hline
$N$   &  $252$  &   $12$  &   $2$   \\    
\hline
\end{tabular}
\label{tab:SavitzkyGolay}
\end{table}

\section{Using Stokes' theorem for calculating~$J_z$}
\label{sec:Stokes}

Instead of using derivatives for calculating $J_z$ (Eq.~\ref{eq:Jz_diff}), one can also use Stokes' theorem:
\begin{equation}
J_z^\mathrm{Int.} = \frac{1}{\mathrm{A_L}} \oint_L \mathbf{B}_\mathrm{hor}\cdot~d\mathbf{L},
\label{eq:Jz_int}
\end{equation}
where the vertical current density is calculated for a pixel in the center of an area $A_L$ outlined by a contour $L$. We calculated the integral by using the composite Simpson's rule over a square, where $L$ is the edge of the square with a side length of 3, 5, or 7 pixels, in accordance to the stencil size when using derivatives.

\begin{figure*}
\centering
\includegraphics[scale=0.65]{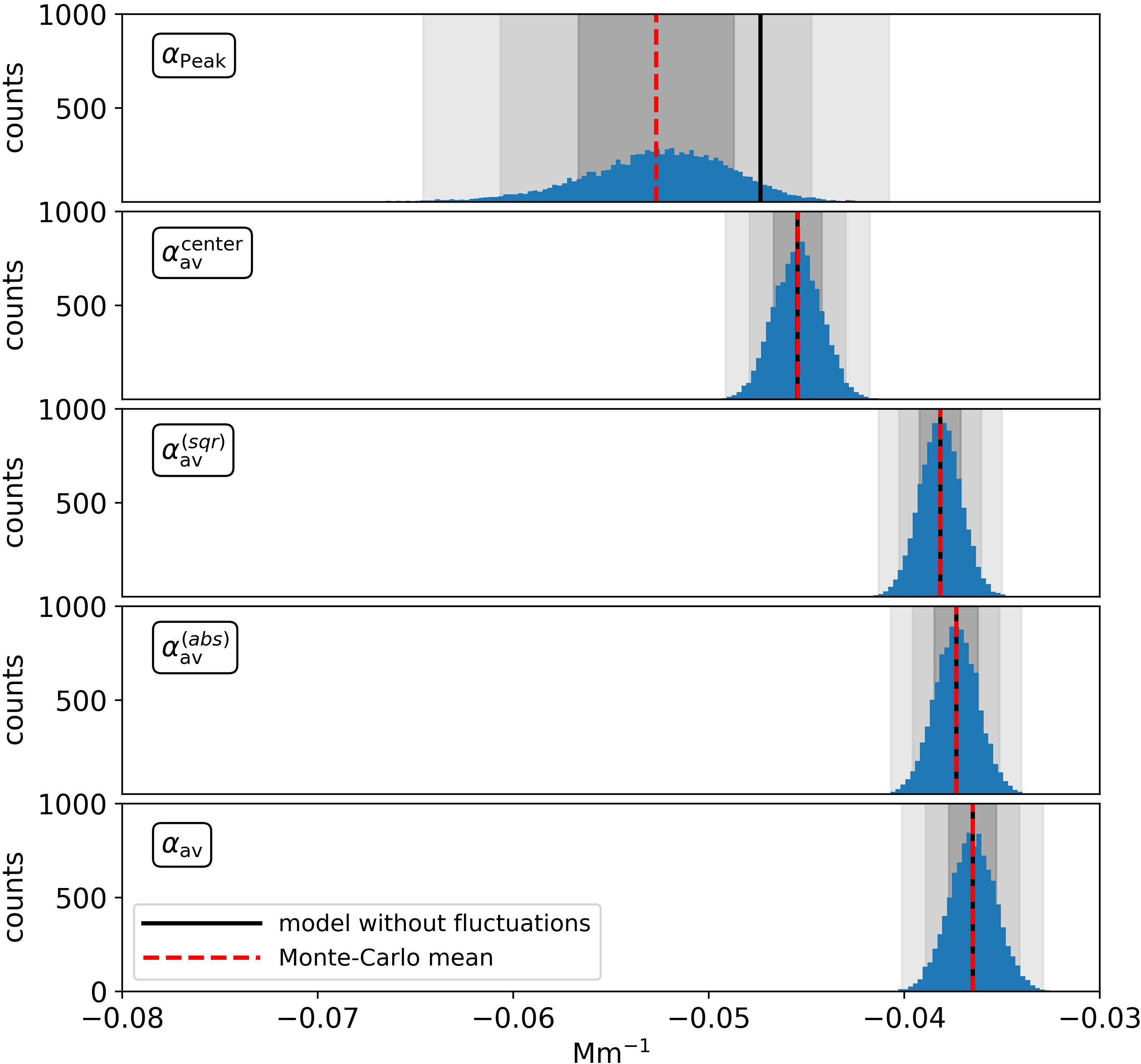}
\caption{Monte-Carlo simulation results for methods based on \aZ when Stokes' theorem is used for calculating $J_z$. The black solid line indicates the reference value of the 
model  that is expected from the model without any fluctuations. The red dashed line shows the mean value from the Monte-Carlo simulations. The dark/light gray shaded areas represent the range of 1, 2, or 3 $\sigma$ around the Monte-Carlo mean.}
\label{fig:twist_hist_int}
\end{figure*}

\begin{figure*}
\centering
\includegraphics[scale=0.65]{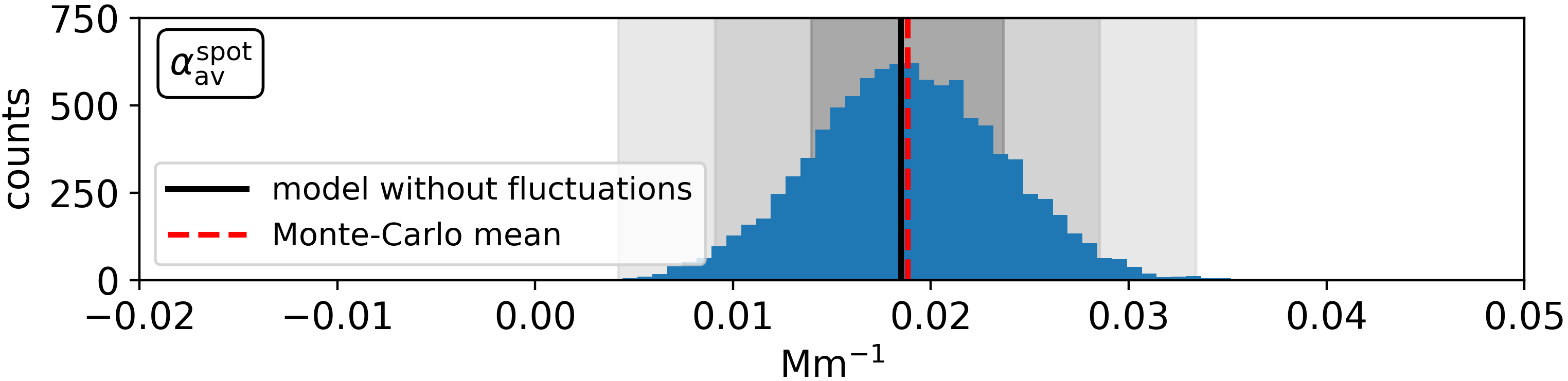}
\caption{Monte-Carlo simulation results when \aZ is calculated with Stokes' theorem and spatially averaged over the whole sunspot. The black solid line indicates the reference value of  the model\\  that is expected from the model without any fluctuations. The red dashed line shows the mean value from the Monte-Carlo simulations. The gray shaded areas represent the range of 1, 2, and 3 $\sigma$ around the mean.}
\label{fig:twist_hist_spot_int}
\end{figure*}

Using Stokes' theorem for calculating \aZ leads on average to slightly lower values compared to the differential form. This effect can be attributed to the integral form using more pixels for the calculations and averaging over an area. The results of Monte-Carlo simulations are shown in Fig.~\ref{fig:twist_hist_int} and Fig.~\ref{fig:twist_hist_spot_int} as well as Table~\ref{tab:results_int}.

\begin{table}[]
\caption{Comparison of the different twist calculation methods between the model reference without any fluctuations of the magnetic field (Model) and the Monte-Carlo simulations (MC Sim.). All parameters were calculated using Stokes' theorem. Values are given in $\mathrm{Mm}^{-1}$.}
\centering
\begin{tabular}{l r r}
\hline\hline
Method & Model & MC Sim.   \\
\hline

$\alpha_\mathrm{Peak}$                              & $-0.047$ & $-0.053 \pm 0.004$ \\
$\alpha_\mathrm{av}^\mathrm{center}$                & $-0.046$ & $-0.046 \pm 0.001$ \\
$\alpha_\mathrm{av}^\mathrm{sqr}$                   & $-0.038$ & $-0.038 \pm 0.001$ \\
$\alpha_\mathrm{av}^\mathrm{abs}$                   & $-0.038$ & $-0.038 \pm 0.001$ \\
$\alpha_\mathrm{av}$                                & $-0.037$ & $-0.037 \pm 0.001$ \\
$\alpha_\mathrm{av}^\mathrm{spot}$                  & $~0.018$ & $~0.019 \pm 0.005$ \\
\hline
\end{tabular}
\label{tab:results_int}
\end{table}

\section{Comparison of different stencil sizes for calculating $J_z$}
\label{sec:boxsize_comparison}

\begin{table*}[]
\caption{Monte-Carlo simulation results for using derivatives to calculate \aZ with stencil sizes of 3, 5, and 7 pixels, when \aZ was evaluated in the umbra of the 
model spot.}
\centering
\begin{tabular}{l c c c}
\hline\hline
 Dif                                   & $3$ & $5$ & $7$ \\
\hline
\aPeak   & $-0.058 \pm 0.006$ & $-0.058 \pm 0.006$ & $-0.058 \pm 0.007$ \\
 \aAvSqr  & $-0.039 \pm 0.001$ & $-0.039 \pm 0.001$ & $-0.039 \pm 0.001$ \\
 \aAvAbs  & $-0.038 \pm 0.001$ & $-0.038 \pm 0.001$ & $-0.038 \pm 0.001$ \\
 \aAv     & $-0.037 \pm 0.001$ & $-0.037 \pm 0.001$ & $-0.037 \pm 0.001$ \\ 
 \hline
\end{tabular}
\label{tab:boxsize_dif}
\end{table*}

\begin{table*}[]
\caption{Monte-Carlo simulation results for using integration to calculate \aZ over square areas with side lengths of 3, 5, and 7 pixels, when \aZ was evaluated in the model spot's umbra.}
\centering
\begin{tabular}{l c c c}
\hline\hline
 Int.                                   & $3$ & $5$ & $7$ \\
\hline
 \aPeak   & $-0.057 \pm 0.006$ & $-0.053 \pm 0.004$ & $-0.050 \pm 0.004$ \\
 \aAvSqr  & $-0.039 \pm 0.001$ & $-0.038 \pm 0.001$ & $-0.038 \pm 0.001$ \\
 \aAvAbs  & $-0.038 \pm 0.001$ & $-0.038 \pm 0.001$ & $-0.037 \pm 0.001$ \\
  \aAv     & $-0.037 \pm 0.001$ & $-0.037 \pm 0.001$ & $-0.036 \pm 0.001$ \\
 \hline
\end{tabular}
\label{tab:boxsize_int}
\end{table*}

We tested stencil sizes of 3, 5, and 7 pixels in Monte-Carlos simulations as described in section~\ref{sec:mc_sim} to determine how many pixels should be considered for calculating the vertical current density $J_z$ and subsequently \aZ. The average twist proxy values and their standard deviation from these simulations are listed in Tables~\ref{tab:boxsize_dif}~and~\ref{tab:boxsize_int}. The different stencil sizes do not noticeably impact the end result when we used the Savitzky-Golay filter. In contrast, we find that with bigger areas the measured twist goes down when Stokes' theorem (see Appendix~\ref{sec:Stokes}) was applied. Similar to the Monte-Carlo simulations described in section~\ref{sec:mc_sim} only the expectation value of the \aPeak method is biased in all test cases. 

\cite{Fursyak2018} tested the effect of differently sized areas for calculating $J_z$ in differential and integral forms using observations from Hinode and HMI. They conclude that integrating over side lengths of five pixels provides the best compromise between smoothing noise but still preserving significant features. As the results from our tests do not show a clear favorite, we chose to calculate $J_z$ with stencil sizes of 5 pixels in accordance to \cite{Fursyak2018}.

\section{Interpretation of \aZ}
\label{sec:a_interpretation}
It is often difficult to interpret \aZ measurements. Ideally we would expect to measure the same \aZ or twist density $q$ value at each location within a uniformly twisted flux tube (which is the case in thin flux tube models). While this is true for the twist density $q$ in our sunspot model without magnetic field fluctuations, \aZ varies with the distance from the center of the spot, where the \aZ profile peaks. 

\cite{Leka2005} used the Gold-Hoyle flux tube model \citep{Gold1960} to show the connection between \aZ and the twist density. This model consists of an axial and azimuthal magnetic field component. \cite{Leka2005} demonstrate that only at the center of the flux tube, where the radial distance from its axis $r$ equals zero, the twist density can be retrieved from \aZ measurements directly. They describe the following relationship between \aZ, the twist density $q$ and the distance from the flux tube axis as:
\begin{equation}
\alpha_z (r) = \frac{2q}{1+q^2r^2}.
\label{eq:alpha_gold}
\end{equation}

We note that in this configuration \aZ goes towards zero when $r$ increases to infinity. This finding led them to propose the \aPeak method.
We can find a similar relationship between \aPeak~ and the twist density $q$ for our empirical sunspot model. Taking the model's equation for \Bt~ (Eq.~\ref{eq:model_Bt}) with $q = \frac{b}{8a_0\sqrt{1+b^2}}$, we get the twist density equation from \cite{Nandy2008}:
\begin{equation}
  B_\theta(r) = q r B_z.
\end{equation}
Together with the model's $B_z$ component,
\begin{equation}
   B_z(r) = B_0 \exp \left[ -\left(\log_\mathrm{e}2\right)\left(\frac{r}{h_0}\right)^2\right],
\end{equation}
we can use derivatives in cylindrical coordinates to calculate~\aZ. We note that $B_r$ in Eq.~\ref{eq:model_Br} does not depend on the azimuthal angle~$\theta$. Then the radial profile of \aZ is:
\begin{equation}
\begin{split}
\alpha_z &= \frac{1}{r B_z} \left[ \frac{\partial (r B_\theta)}{\partial r} - \xcancel{\frac{\partial B_r}{\partial \theta}} \right] = \\
    & = \frac{1}{r B_z} \frac{\partial}{\partial r} \left\{ q B_0 r^2 \exp 
    \left[-\left(\log_\mathrm{e}2\right)\left(\frac{r}{h_0}\right)^2\right]\right\} = \\
    &= 2 q \left[1 - \left(\log_\mathrm{e}2\right)\left(\frac{r}{h_0}\right)^2\right]
\end{split}.
\label{eq:alpha_cameron}
\end{equation}
We get a maximum value of \aZ (\aPeak) at the center of the flux tube, i.e. $r=0$. Therefore, we find the same relationship of $\alpha_\mathrm{Peak} = 2q$ as \cite{Leka2005}. \\

In our empirical sunspot model \aZ changes sign at a certain distance from the suncenter of the spot. We can determine where alpha changes sign by setting \aZ in Eq.~\ref{eq:alpha_cameron} equal to zero to get
\begin{equation}
r_{\alpha_z=0} = \frac{h_0}{\sqrt{\log_\mathrm{e}2}} \approx 1.2~h_0.
\label{eq:alpha_equals_zero}
\end{equation}
The parameter $h_0$ describes the distance of the umbra--penumbra boundary measured from the model center of the spot. Disregarding fluctuations of the magnetic field, within the umbra the sign of the twist and \aZ are the same, according to our model.

Since $\alpha_z = J_z/B_z$, such a sign reversal can only happen in our sunspot model with positive polarity, if the sign of the vertical current density $J_z$ changes. Such rings of return currents at the umbra--penumbra boundary of observed sunspots are reported in literature \citep[e.g.,][]{Tiwari2009}. We note that positive and negative vertical currents are balanced in our model.

\end{document}